\documentclass[twocolumn,aps,prl,superscriptaddress, amsmath,amssymb,longbibliography,nofootinbib]{revtex4-2}

\usepackage{graphicx}
\usepackage{siunitx}	
\usepackage{amssymb}
\usepackage{datetime}
\usepackage[T1]{fontenc}
\usepackage{booktabs}
\usepackage{hyperref}
\usepackage{xcolor}
\hypersetup{
    colorlinks,
    linkcolor={red!50!black},
    citecolor={blue!50!black},
    urlcolor={blue!80!black}
}

\usepackage{todonotes}

\newcommand{\icm}{\ensuremath{~\textrm{cm}^{-1}} }

\begin{document}

%
%

\preprint{ver 0.7 \today\ \currenttime}

\title{Pressure-induced formation of cubic lutetium hydrides derived from trigonal LuH$_3$}

\author{Owen Moulding\footnote{owen.moulding@neel.cnrs.fr}}
\affiliation{Institut N\'{e}el CNRS/UGA UPR2940, 25 Avenue des Martyrs, 38042 Grenoble, France}
\author{Samuel Gallego-Parra}
\affiliation{European Synchrotron Radiation Facility, 71 Avenue des Martyrs, 38000 Grenoble, France}
\author{Yingzheng Gao}
\affiliation{Institut N\'{e}el CNRS/UGA UPR2940, 25 Avenue des Martyrs, 38042 Grenoble, France}
\author{Pierre Toulemonde}
\affiliation{Institut N\'{e}el CNRS/UGA UPR2940, 25 Avenue des Martyrs, 38042 Grenoble, France}
\author{Gaston Garbarino}
\affiliation{European Synchrotron Radiation Facility, 71 Avenue des Martyrs, 38000 Grenoble, France}
\author{Patricia De Rango}
\affiliation{Institut N\'{e}el CNRS/UGA UPR2940, 25 Avenue des Martyrs, 38042 Grenoble, France}
\author{S\'{e}bastien Pairis}
\affiliation{Institut N\'{e}el CNRS/UGA UPR2940, 25 Avenue des Martyrs, 38042 Grenoble, France}
\author{ Pierre Giroux}
\affiliation{Institut N\'{e}el CNRS/UGA UPR2940, 25 Avenue des Martyrs, 38042 Grenoble, France}
\author{Marie-Aude M\'{e}asson\footnote{marie-aude.measson@neel.cnrs.fr}}
\affiliation{Institut N\'{e}el CNRS/UGA UPR2940, 25 Avenue des Martyrs, 38042 Grenoble, France}

\date{\today}

\begin{abstract}

In recent years, there has been a fervent search for room-temperature superconductivity within the binary hydrides. However, as the number of untested compounds dwindled, it became natural to begin searching within the ternary hydrides. This led to the controversial discovery of room-temperature superconductivity at only 1\,GPa in nitrogen-doped lutetium hydride [Dasenbrock-Gammon \textit{et al.}, Nature \textbf{615}, 244 (2023)] and consequently provided much impetus for the synthesis of nitrogen-based ternary hydrides. Here, we report the synthesis of stable trigonal LuH$_3$ by hydrogenating pure lutetium which was subsequently pressurised to $\sim$2\,GPa in a dilute-N$_2$/He-rich pressure medium. Raman spectroscopy and x-ray diffraction were used to characterise the structures throughout. After depressurising, energy-dispersive and wavelength-dispersive X-ray spectroscopies characterised the final compound. Though our compound under pressure exhibits similar structural behaviour to the Dasenbrock-Gammon \textit{et al.} sample, we do not observe any nitrogen within the structure of the recovered sample at ambient pressure. We observe two cubic structures under pressure that simultaneously explain the X-ray diffraction and Raman spectra observed: the first corresponds well to $Fm\overline{3}m$ LuH$_{2+x}$, whilst the latter is an $Ia\overline{3}$-type structure.


\end{abstract}

\maketitle

\section{Introduction}

The holy grail of room-temperature superconductivity has been a long sought-after quest, ever since the initial predictions of superconductivity in metallic hydrogen by Ashcroft in 1968 \cite{Ashcroft1968} and shortly after the publication of BCS theory in 1957 \cite{Bardeen1957, Bardeen1957a}. Though not pure hydrogen, many examples of high-temperature superconductivity have been realised in recent years; these have reliably shattered high-critical-temperature (high-$T_c$) records with each new discovery. A notable example was SH$_3$ with a $T_c$ of 203\,K at 155\,GPa \cite{Drozdov2015}, as it provided tangible promise for the field. Subsequent examples continued to push the threshold with the discovery of superconductivity in YH$_9$ and LaH$_{10}$ at 243 and 260\,K respectively both at approximately 200\,GPa \cite{Kong2021, Somayazulu2019, Drozdov2019}. Clearly these superconducting states require extremely high pressures that few groups are able to reach, and this has been the primary technical challenge to overcome.

Hence why the claim of room-temperature superconductivity at 294\,K in nitrogen-doped (N-doped) lutetium hydride at such a low pressure of 1\,GPa \cite{Dasenbrock-Gammon2023} has drawn so much attention. Not only is it a new record $T_c$ for superconductivity, but also it brings superconductivity into the domain of practicably achievable at near-ambient conditions. Furthermore, the samples are said to be metastable at ambient pressure which further adds to the wishful properties of such a material. In such a short period of time, an impressive number of groups have already tried to replicate the results, both theoretically and experimentally \cite{Shan2023, Huo2023, Li2023, Sun2023, Xie2023, Ming2023a, Xing12023, Zhang2023, Hilleke2023}, though a corroborative synthesis remains elusive. Even $Nature$ has recently published an article entitled ``Absence of near-ambient superconductivity in LuH$_{2+x}$N$_y$'' by Ming \textit{et al.} \cite{Ming2023a} in direct contention with the original $Nature$ publication \cite{Dasenbrock-Gammon2023}, which goes to show how controversial this discovery has been.

N-doped lutetium hydride represents another step into the domain of ternary compounds following the exhaustive hunt for binary hydride room-temperature superconductors. This new domain is much larger and therefore more daunting to explore, so theoretical predictions are vital to guide experimental works, and they have already yielded several candidate compounds: Li$_2$MgH$_{16}$ \cite{Sun2019_1, Wang2020_1}, YCaH$_{12}$ \cite{Xie2019_1, Liang2019_1}, ScYH$_6$ \cite{Wei2021_1}; and also the LaH$_{10}$-like clathrate boronitrides La(BN)$_5$ and Y(BN)$_5$ \cite{Ding2022_1}. Calculations optimising superconductivity via doping have also shown that nitrogen from ammonia borane may affect the superconducting properties of LaH$_{10}$ \cite{Wang2020_1, Ge2021_1, Cataldo2022_1}. Experimentally, the most notable confirmed example of a ternary hydride comes from $Fm\overline{3}m$-(La,Y)H$_{10}$ with a superconducting temperature of 253\,K at 183\,GPa \cite{Semenok2021_1}. Beyond this, synthesising high-quality, high-$T_c$ ternary compounds under extreme pressures remains rare, thus efforts that characterise this phase space in such extreme environments are vital for the field.

In order to synthesise N-doped lutetium hydride, Dasenbrock-Gammon \textit{et al.} \cite{Dasenbrock-Gammon2023} and Cai \textit{et al.} \cite{Cai2023} used pure lutetium with a H$_2$/N$_2$ gas mixture, whereas other experimental papers started from pure lutetium and NH$_4$Cl and CaH$_2$ precursors \cite{Xing12023, Ming2023a} which decompose to provide the required N$_2$ and H$_2$. Here we choose another process, by first synthesising pure LuH$_3$ and then loading the diamond anvil cell (DAC) with a mixture of dilute N$_2$ and helium. We then methodically characterise the obtained compound with Raman spectroscopy and x-ray diffraction (XRD) at each step, and by x-ray energy-dispersive-spectroscopy (EDS) and wavelength-dispersive-spectroscopy (WDS) at ambient pressure.

\section{Methods}

\subsection{Experimental Methods}

In total we prepared three DACs with thin samples of presynthesised LuH$_3$. Prior to synthesis, polished lutetium metal was characterised by EDS and oxygen and tantalum were observed in small quantities. The LuH$_3$ was then synthesised by hydrogen absorption using the Sievert method by heating for 18\,hours at 200\,$^{\circ}$C in 4\,MPa of H$_2$ gas; further synthesis details are provided in the Supplementary Material (SM), Sec. S1 \cite{SI}. All samples came from this synthesis and were distributed amongst the three DACs. The first DAC (DAC1) was loaded with a mixture of nitrogen and helium, where we estimate that the quantity of N$_2$ in the pressure chamber was 4\,nmol whilst the quantity of LuH$_3$ was 11\,nmol. The other two DACs (DAC2 and DAC3) were loaded with nitrogen: DAC2 was loaded with a gas loader, whereas DAC3 was cryogenically loaded with liquid nitrogen. Amongst the DACs, only the sample within DAC1 showed structural and chemical transformations under pressure which are discussed in the main text of this paper. The other DACs and further details are discussed in the SM \cite{SI}. A ruby ball (for pressure measurement) and a piece of silicon (for optimising the Raman signal) were also placed inside the pressure chamber. DAC1 was sealed at 1.9\,GPa and characterised by Raman spectroscopy and XRD. Though the sample was eventually heated to 65\,°C at 1.9\,GPa, the main text only presents data prior to heating, as heating had no effect on the structural properties. 

The XRD study was performed on the European Synchrotron Radiation Facility (ESRF) ID15B beamline with $\lambda$=0.411\,\AA~at 300\,K. Polarised Raman scattering was performed in quasi-backscattering geometry at 300\,K with an incident laser line at 532\,nm from a solid-state laser. The scattered light was analysed by a single-grating spectrometer and a triple-grating subtractive spectrometer; both were equipped with liquid-nitrogen-cooled CCD detectors. We measured the Raman signal of pure LuH$_3$ just before loading in the DAC, after loading at 1.9\,GPa, before and after heating, and finally after returning to ambient pressure. After depressurising, we analysed the composition of the sample with EDS and WDS whilst primarily searching for nitrogen.


\section{Experimental results}

\subsection{Imaging of the sample}

\begin{figure}[h!]
\centering
\includegraphics[width=1\linewidth]{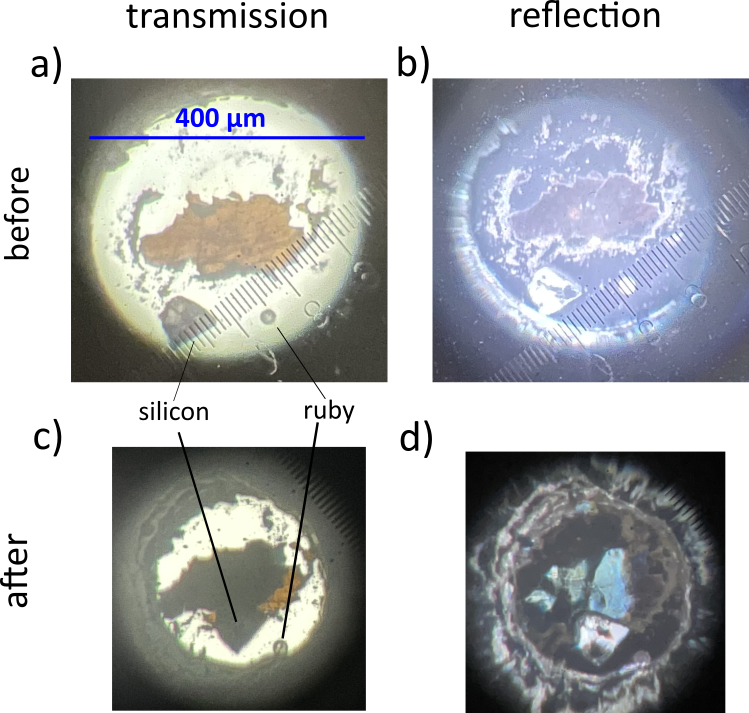}
\caption{White light images of the sample before [(a) and (b)] and after [(c) and (d)] loading at 1.9\,GPa. Transmission images are shown in [(a) and (c)] and reflection images are shown in [(b) and (d)].}
\label{figImage}
\end{figure}

The colour change from blue at ambient pressure to red at high pressure has been actively discussed in the literature \cite{Dasenbrock-Gammon2023,Shan2023,Xing12023,Zhang2023}. Images of our sample in DAC1 before (300\,K, 1\,bar) and after (300\,K, 1.9\,GPa) loading are presented in Fig.~\ref{figImage}. A white light was used to illuminate the sample in reflection and in transmission. Our LuH$_3$ sample appears translucent with a red colour at 1\,bar and seems to become opaque at high pressure; however, this could be due to the majority of the sample rising up off of the diamond during loading. After loading with the mixture of He/N$_2$ and pressurising to 1.9\,GPa, the surface became reflective and blue. In Fig. \ref{figImage}c, we can also see a red region which remained flat against the diamond which was also characterised and is discussed in Sec. S2 of the SM \cite{SI}. 

\subsection{X-ray diffraction}

\begin{figure}[h!]
\centering
\includegraphics[width=1\linewidth]{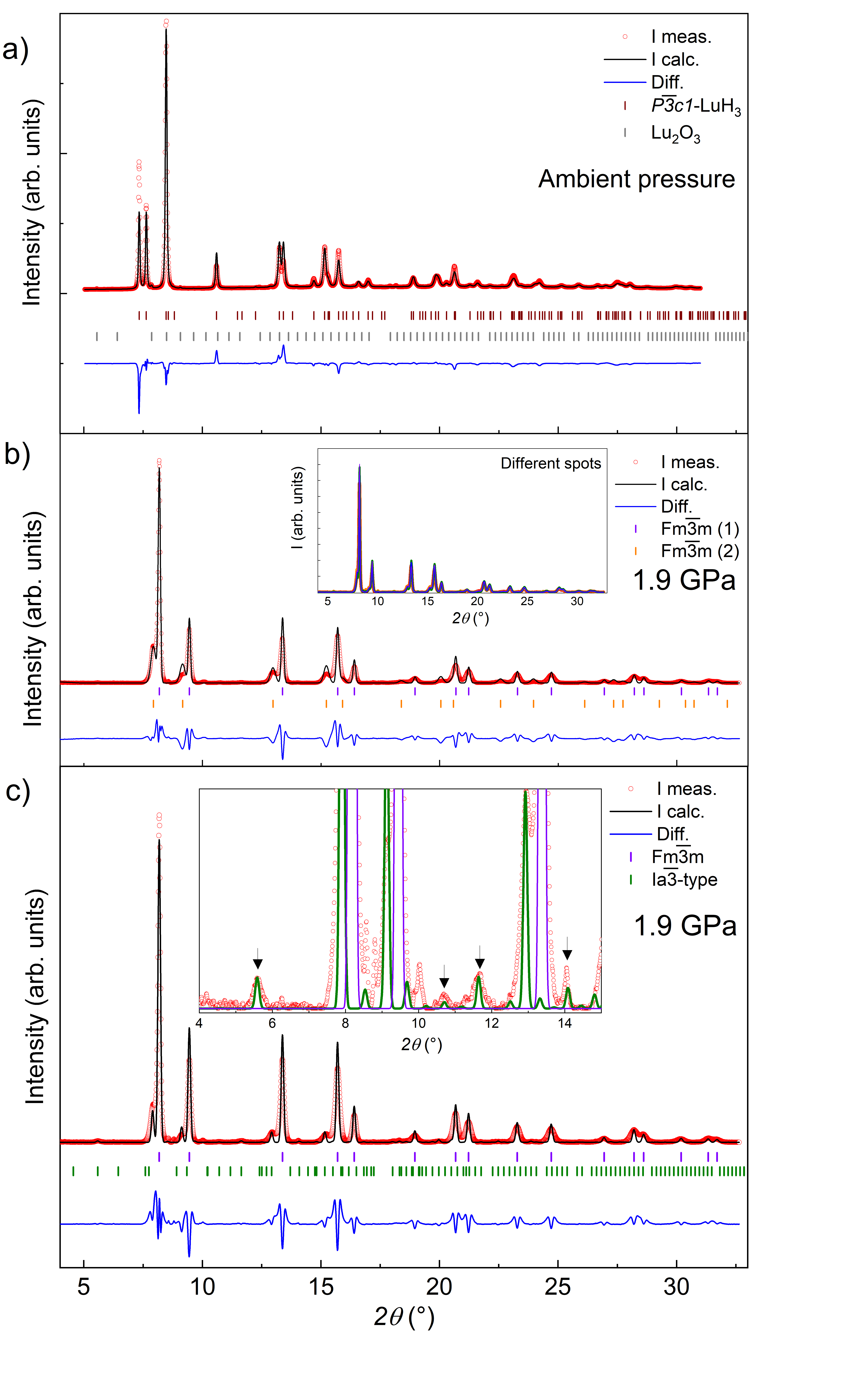}
\caption{Rietveld refinements of the patterns measured at the ESRF (beamline ID15B, $\lambda$\,=\,0.411\,\AA) at 300\,K. (a) The trigonal LuH$_3$ sample at ambient pressure. (b) The high pressure compound at 1.9\,GPa and fitted with two $Fm\overline{3}m$  structures, structures 1 and 2. Inset: patterns measured on five different spots. (c) The high-pressure compound at 1.9\,GPa and fitted with one $Fm\overline{3}m$ structure and one $Ia\overline{3}$-type structure. Inset: zoom of some of the weak reflections fitted by the $Ia\overline{3}$-type structure (cf. arrows). Diff., difference between measured and calculated values.}
\label{figXRD2}
\end{figure}

The Rietveld fit of the XRD pattern measured on the trihydride in ambient conditions is shown in Fig. \ref{figXRD2}(a), and we determine the structure to be trigonal $P\overline{3}c1$ with lattice parameters of $a$\,=\,6.173(1)\,\AA~ and $c$\,=\,6.424(1)\,\AA. The lanthanide trihydrides tend to adopt either this trigonal structure or a hexagonal $P6_3/mmc$ structure (the higher-symmetry parent group) \cite{Kong2012}. Previously, Tkacz and Palasyuk \cite{Tkacz2007} determined that LuH$_3$ is hexagonal with $a$\,=\,3.57\,\AA~ and $c$\,=\,6.41\,\AA~ at ambient conditions. However, previous measurements had already shown that the structure is trigonal with lattice parameters of $a$\,=\,6.16\,\AA~ and $c$\,=\,6.44\,\AA~\cite{Mansmann1964} which are similar to our values. Furthermore, recent calculations by Dangi\'c \textit{et al.} predict that the trigonal structure should be more stable than the hexagonal structure in this pressure range \cite{Dangic2023}. Finally, the hexagonal structure would also be inconsistent with the Raman spectra we measured due to having too few excitations, as shown in Table SIV of Sec. S5 in the SM \cite{SI}. Overall we conclude that our starting LuH$_3$ adopts a trigonal $P\overline{3}c1$ structure in ambient conditions.

With regard to impurities within our sample, from the Rietveld fit we determine that the sample is primarily LuH$_3$ at 96.9(1)$\%$, and the rest was identified to be Lu$_2$O$_3$. The Lu$_2$O$_3$ is likely to originate from deposits on the lutetium surface that were not removed by polishing before hydrogenation. The space group of Lu$_2$O$_3$ is $Ia\overline{3}$ and the refined lattice parameter is 10.380(8) \AA~ in agreement with the literature \cite{Lin2010, Jiang2010a}. We also show that the percentage of Lu$_2$O$_3$ stays constant for 6 months with the sample exposed directly to air (Sec. S2 of the SM \cite{SI}); so the sample is stable with respect to oxidation within this time scale. The EDS measurements showed that a small quantity of tantalum was present in the starting lutetium; however, there are no signatures of tantalum or tantalum hydride in the XRD spectra.

XRD patterns from the loaded sample at 1.9\,GPa are shown in Fig. \ref{figXRD2}(b). They were measured in five different spots with sizes of 4 x 3\,$\mu$m and separated by 20\,$\mu$m in a cross-shape. The results on the different spots are remarkably similar and indicate that the sample is homogeneous in this region [see inset of Fig. \ref{figXRD2}(b)]. By comparing the XRD patterns, the transformation to a new phase is clear. In their paper, Dasenbrock-Gammon \textit{et al.} determine the synthesised ambient pressure sample to consist of two distinct $Fm\overline{3}m$ phases \cite{Dasenbrock-Gammon2023}: the majority LuH$_{3-\delta}$N$_\epsilon$ ``A'' phase (92.25\,$\%$ of the sample) has a lattice parameter of $a_A$\,=\,5.0298(4)~\AA, whilst the lattice parameter of the minority LuN$_{1-\delta}$H$_\epsilon$ ``B'' phase (7.29\,$\%$) is $a_B$\,=\,4.7529(9)~\AA~\cite{Dasenbrock-Gammon2023}. Under pressure at 1.9\,GPa, we obtain similar XRD patterns that can be reasonably well-described by two $Fm\overline{3}m$ phases. Our majority phase ($\approx$\,60$\%$) has a lattice parameter of $a_1$=4.990(6)\,\AA, whilst our minority phase ($\approx$\,40$\%$) has a lattice parameter of $a_2$=5.145(2)\,\AA. We note that our majority phase is the one with the smaller lattice parameter, but more disconcertingly we notice that the lattice parameters of both of our phases are larger than those of Dasenbrock-Gammon \textit{et al.} despite our compound being under pressure. A tempting explanation might rely on the synthesis process which starting from pure LuH$_3$ would tend to produce compounds with higher hydrogen content that are closer to the trihydride with an expanded lattice.

Interestingly, after pressurisation there are some small reflections that cannot be described by the refinement using two $Fm\overline{3}m$ phases. Moreover, there is a clear inconsistency between the two $Fm\overline{3}m$ phases and the Raman spectra, as shall be discussed in more detail later. This leads us to reconsider the structural composition, and our analysis is in favour of one $Fm\overline{3}m$ structure and one $Ia\overline{3}$ structure.\\ 

Indeed, Fig. \ref{figXRD2}(c) shows that the small reflections can be better explained by refining the XRD data at 1.9\,GPa with one $Fm\overline{3}m$ structure and one $Ia\overline{3}$ structure. From this refinement, we obtained lattice parameters of 4.99(3)\,\AA~ and 10.329(3)\,\AA~ for the $Fm\overline{3}m$ and $Ia\overline{3}$ structures respectively. The lattice parameter of the $Fm\overline{3}m$ structure remains the same within error as that of the previous refinement using two $Fm\overline{3}m$ structures. Here we exclude the presence of $Fm\overline{3}m$ LuH$_{3}$, since this phase was only observed previously above 12\,GPa \cite{Tkacz2007}, far beyond our measured pressure range. However, other $Fm\overline{3}m$ compounds remain possible and shall be discussed later.\\ 

Regarding the $Ia\overline{3}$ phase, we notice that it is similar to the second $Fm\overline{3}m$ structure but with an approximate doubling of the lattice parameter (2$a_2$, eight times the volume) and a slightly lower symmetry. Though the $Ia\overline{3}$-type structure is similar to the $Fm\overline{3}m$ structure, the lutetium atoms occupy different Wyckoff positions within the lattice: namely the $8b$ and $24d$ sites. The $8b$ site is highly symmetric, (1/4,\,1/4,\,1/4), whilst the $24d$ site is described by ($x$,\,0,\,1/4) where $x$ was determined to be approximately 0.975(8). This small difference from unity is indicative of a slight distortion in the lutetium sublattice relative to the global cubic symmetry. The occupation of the $24d$ site also has ramifications for the Raman activity as it provides eight additional phonons, whereas the $8b$ site does not provide any. This shall be discussed further in later sections.\\ 

Even though the $Ia\overline{3}$ phase is reminiscent of Lu$_2$O$_3$, we state that it is not the same compound. Firstly, the lattice parameter is smaller than the value of 10.357\,\AA~ for Lu$_2$O$_3$ at 1.9\,GPa, which was determined from the volume dependence of Ref. \cite{Jiang2010a}. Secondly, since the $Ia\overline{3}$ compound is recoverable (though metastable on the timescale of days as shown in Sec. S3 of the SM), we determine that the ambient pressure lattice parameter is 10.41(1)\,\AA~ (see Sec. S3 of the SM) which is larger than the ambient pressure value for Lu$_2$O$_3$ of 10.38\,\AA ~\cite{Jiang2010a}. Together, these lattice parameters at ambient and high pressure indicate that the $Ia\overline{3}$ phase has a larger compressibility than Lu$_2$O$_3$ which further distinguishes them as separate compounds. Finally, the Raman spectrum, as shown in the next section, does not contain the expected main Raman mode of Lu$_2$O$_3$. Therefore, we conclude that the high-pressure sample of DAC1 does not contain two $Fm\overline{3}m$ phases, but in fact one $Fm\overline{3}m$ phase and one $Ia\overline{3}$ phase that we shall label as an $Ia\overline{3}$-type phase henceforth.

\subsection{Raman spectroscopy}

We first recall the nature of the $\Gamma$-point phonons expected in the various space groups under consideration (see Sec. S5 of the SM for more space groups \cite{SI}). From the literature on LuH$_3$ (and YH$_3$), the crystal structure could correspond to $Fm\overline{3}m$ or $P\overline{3}c1$ \cite{Kong2012, Palasyuk2005,Kierey2001}. We expect a total of $5A_{1g}\oplus12E_{g}$ Raman active phonon modes in the trigonal $P\overline{3}c1$ phase, and a single Raman-active $T_{2g}$ mode in the $Fm\overline{3}m$ structure, as stated in Table \ref{tab}. The $T_{2g}$ mode is associated with the displacement of the hydrogen atoms occupying the $8c$ Wyckoff sites and is also expected to appear in $Fm\overline{3}m$ LuH$_{2}$ and $Fm\overline{3}m$ LuH$_{2+x}$. Here we note that the $Fm\overline{3}m$ LuH$_2$ and LuH$_3$ are related by the partial and continuous occupation of the octahedral $4b$ sites which results in the formation of LuH$_{2+x}$. Spectroscopically and as shown in Table \ref{tab}, $Fm\overline{3}m$ LuH$_{3}$ and LuH$_{2+x}$ behave very similarly, whilst $Fm\overline{3}m$ LuH$_2$ lacks a $T_{1u}$ mode since the $4b$ site is completely unoccupied.

\begin{table}[]
\footnotesize
\begin{tabular}{c|c|c|c|c|c|c}
\hline \hline
       Space group & Lu & H$^{1}$ & H$^{2}$ & H$^{3}$ & IR-active & R-active    \\ \hline
$Fm\overline{3}m$ (LuH$_3$ \cite{Sun2023}) & $4a$ & $8c$ & $4b$ & - & $2T_{1u}$ & $1T_{2g}$ \\ \hline
$Fm\overline{3}m$ (LuH$_{2+x}$) & $4a$ & $8c$ & $4b$ & - & $2T_{1u}$ & $1T_{2g}$ \\ \hline
$Fm\overline{3}m$ (LuH$_2$ \cite{Sun2023}) & $4a$ & $8c$ & - & - & $1T_{1u}$ & $1T_{2g}$ \\ \hline
$P\overline{3}c1$ (YH$_3$ \cite{Kierey2001}) & $6f$ & $2a$ & $4d$ & $12g$ & $6A_{2u}+11E_{u}$ & $5A_{1g}+12E_{g}$  \\ \hline
\addlinespace
\addlinespace
\hline
Space group & Lu$^{1}$ & Lu$^{2}$ & H$^{1}$ & H$^{2}$ & IR-active & R-active    \\ \hline
$Ia\overline{3}$ ($Ia\overline{3}$-type) & $8b$ & $24d$ & - & - & $7T_{u}$ & $1A_{g}$+$2E_g$
\\  & & & & & &+$5T_{g}$ \\ \hline\hline
\end{tabular}
\caption{The total number of optical infrared-active (IR-active) and Raman-active (R-active) modes for the given space groups with the occupied Wyckoff positions stated for various compounds. }
\label{tab}
\end{table}

  

\begin{figure}[h!]
\centering
\includegraphics[width=1\linewidth]{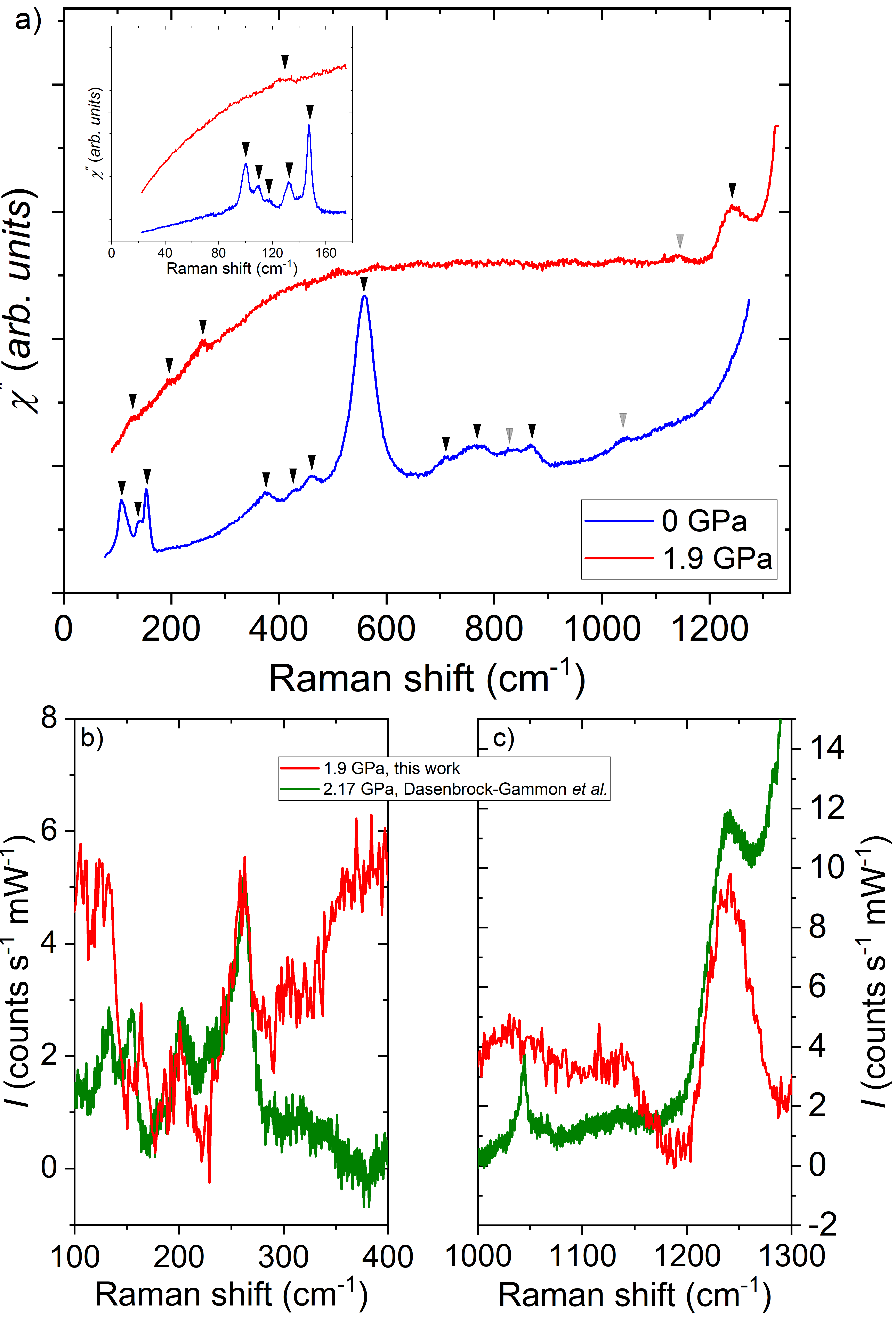}
\caption{ (a) Raman spectra of trigonal LuH$_3$ at ambient pressure (blue) and a high-pressure sample at 1.9\,GPa (red). The inset shows low-energy triple-stage data. (b) and (c) show our data scaled on the Dasenbrock-Gammon \textit{et al.} data at $\sim$2\,GPa  \cite{Dasenbrock-Gammon2023}. We scale on the peak at 260\,\si{cm^{-1}} after a background correction which aids the comparison. The scaling in (b) is the same as in (c).}
\label{figRaman2}
\end{figure}


Wide-range Raman spectra on the ambient pressure trigonal LuH$_3$ and the high-pressure sample are shown in Fig. \ref{figRaman2}(a). For the ambient pressure trigonal phase, we observe at least 12 features that are marked by black arrows. This is close to the 17 phonon modes expected for the trigonal $P\overline{3}c1$ structure and supports our XRD analysis. Importantly, the number of modes far exceeds the four phonon modes predicted for the alternative hexagonal $P6_3/mmc$ structure (see Sec. S5 of the SM); so we can conclusively exclude it as a viable structure. As we increase the pressure, we clearly observe the disappearance of all the phonons observed associated with the trigonal phase which is indicative of a structural transition. We also observe a large increase in the background by a factor of $\sim$\,10, though we cannot conclude whether it is intrinsic or due to the angle of the sample as compared with the diamond. Most notably, we observe two peaks at high pressure that consistently appear at approximately 1240 and 260\,\si{cm^{-1}} which were not present at ambient pressure.\\

At energies below 260\,\si{cm^{-1}} we observe other features, most notably three weak excitations at 202, 164, and 128\,\si{cm^{-1}}. As shown in Fig. \ref{figRaman2}(b), these are similar to not only those observed by Dasenbrock-Gammon \textit{et al.} \cite{Dasenbrock-Gammon2023} but also those osberved by Xing \textit{et al.} \cite{Xing12023}, who ascribed them to vibrational modes of $Fm\overline{3}m$ compounds. However, the number of Raman modes is inconsistent with two $Fm\overline{3}m$ structures, as we only expect one $T_{2g}$ mode for each phase. Furthermore, we do not expect the lower-symmetry Wyckoff sites (e.g. $24e$, $32f$, etc.) to become occupied since hydrogen concentrations above thee H atoms per Lu atom have not been observed at these pressures. Herein lies the contradiction with these previous analyses: two $Fm\overline{3}m$ structures cannot explain the number of phonon modes observed here and previously \cite{Dasenbrock-Gammon2023,Xing12023}. On the other hand, a distortion to a $Ia\overline{3}$-type phase with lutetium atoms on the $24d$ Wyckoff sites provides $1A_g\oplus2E_g\oplus5T_g$ phonon modes, and since the lutetium atoms are heavy, these phonon modes would be at low energy. Thus the $Ia\overline{3}$-type phase could provide the required modes at low energy that were observed by us and others \cite{Xing12023, Dasenbrock-Gammon2023}.

\section{Discussion}


To summarise the results, from the XRD we have observed a biphasic mixture of cubic $Fm\overline{3}m$ and cubic $Ia\overline{3}$ by accounting for the numerous weak reflections. These weak reflections are not described by two $Fm\overline{3}m$ structures. From the Raman spectroscopy, we observe one strong mode at 1240\,\si{cm^{-1}} and several weak modes at and below 260\,\si{cm^{-1}}. The number of modes cannot be explained by two $Fm\overline{3}m$ structures, whereas the $Ia\overline{3}$ structure can in principle provide many modes at low energy. As clearly stated by Hilleke \textit{et al.} \cite{Hilleke2023}, from the XRD results the identified sublattices of lutetium atoms (fcc for an $Fm\overline{3}m$ structure and bcc for an $Ia\overline{3}$ structure) provides a constraint about which we should search but it does not necessarily describe the entire structure. Now we shall discuss the possible origin of these structures, and whether or not known compounds can explain the data.

Firstly, we shall address the contaminants which include Lu$_2$O$_3$, pure tantalum, TaH$_{1-x}$, and the van der Waals solid He(N$_2$)$_{11}$ \cite{Vos1992}. This last compound forms beyond the pressure range of interest (above 9\,GPa) and the stoichiometry of the pressure medium is vastly different from that of the compound, so we do not think that it is present. We have already shown that the Lu$_2$O$_3$ impurities are minor in our XRD pattern at ambient pressure ($\approx$\,3$\%$), so we do not expect a large effect from their presence. Furthermore, we do not see any Raman signature of this phase. Indeed, the most intense Raman-active mode of Lu$_2$O$_3$ is observed at 390\,\si{cm^{-1}} at ambient pressure (shown in Sec. S3 of the SM \cite{SI}) and hardens slightly up to 400\,\si{cm^{-1}} at 2\,GPa \cite{Jiang2010a}. However, there is no indication of this mode in any of the locations measured. Therefore we eliminate Lu$_2$O$_3$ as being responsible for the XRD pattern and Raman-active modes, at either ambient or high pressure. Though the quantity is small ($\approx$\,1\%), pure tantalum and TaH$_{1-x}$ could potentially be present. Pure tantalum forms an $Im\overline{3}m$ structure 
\cite{Kuzovnikov2017}, whereas TaH$_{1-x}$ forms an $I\overline{4}m2$ structure
~ \cite{Kuzovnikov2021}. Neither structure can explain the XRD reflections, and so we also eliminate pure tantalum and TaH$_{1-x}$ from consideration.

One should also consider intercalation effects from the pressure medium itself. Previous measurements have shown that helium can occupy interstitial voids and change the structural properties of materials under pressure \cite{Sato2011a,Merlen2005, Merlen2006,schirber1995,yagi2007}. This effect seems confined to network-forming structures \cite{Sato2011a} or to materials possessing large voids such as single-wall carbon nanotubes \cite{Merlen2005, Merlen2006}, fullerenes \cite{schirber1995}, or clathrates \cite{yagi2007}. However, neither trigonal, $Fm\overline{3}m$, nor $Ia\overline{3}$-type phases form these types of structures, and so we do not expect such helium intercalation; see Sec. S2 of the SM for further discussion. Nor would we expect an intercalation effect from N$_2$ molecules due to their much larger size.

We will now compare our XRD and Raman results with the known phases in the Lu-H-N landscape at room temperature and $\sim$\,2\,GPa. These consist of pure N$_2$ phases, $Fm\overline{3}m$ ammonia (NH$_3$) \cite{Gauthier1986,Gauthier1988}, fcc rock-salt LuN (RS-LuN; NaCl-type $B_1$, $Fm\overline{3}m$), fcc zinc-blende LuN (ZB-LuN; ZnS-type $B_3$, $F\overline{4}3m$), hexagonal  LuH$_\delta$ ($P6_3/mmc$), and fcc LuH$_2$ (CaF$_2$-type, $Fm\overline{3}m$).

At room temperature and 2\,GPa, pure N$_2$ may form either a fluid or a solid $\beta$ phase. The $\beta$-phase crystallises in a $P6_3/mmc$ structure \cite{Ouillon1990, Buchsbaum1984}, and a single mode is expected at $\sim$~2330\,\si{cm^{-1}}, which we observe as a narrow peak in this range of energy. N$_2$ gas has not only a similar vibron mode at high energy but also other peaks at low energy below 150\,\si{cm^{-1}} \cite{Ohno2021}. Some of the modes that we measured might originate from N$_2$ gas, but not the ones at 195 and 166\,\si{cm^{-1}} or our dominant modes at 1240 or 260\,\si{cm^{-1}}. 

Ammonia could in principle form if hydrogen liberated from the trigonal LuH$_3$ lattice reacted with nitrogen instead of being replaced by it. At 2\,GPa and ambient temperature, ammonia is expected to form a $Fm\overline{3}m$ structure which should only possess one Raman-active mode \cite{Gauthier1988, Ninet2006}. Ammonia is unlikely to be detected by XRD due to the weak signal from the light atoms contrasted against the large contribution from the massive lutetium atoms, therefore it is unlikely that any of the refined cubic phases could originate from it. Raman scattering under pressure shows that only modes at energies higher than 3100\,\si{cm^{-1}} are observed in this phase \cite{Gauthier1988}. So we exclude ammonia from being responsible for the Raman modes we measure at 1.9\,GPa.


The primary potential nitride compound is $Fm\overline{3}m$ RS-LuN which has a lattice parameter of $a$\,=\,4.7563(4)\,\AA~ at ambient conditions \cite{Suehiro2004}. Therefore this cannot explain either of the two cubic phases observed by XRD, as the lattice parameter will only continue to shrink under pressure and it is already smaller than both of the lattice parameters measured. Furthermore, RS-LuN is in principle Raman inactive since only the $4a$ and $4b$ Wyckoff sites are occupied. Despite this, a strong excitation was observed previously at 582\,\si{cm^{-1}} and was ascribed to strong disorder \cite{Granville2009}. Regardless, we do not observe this mode. We also note that the synthesis of RS-LuN is challenging and previously required heating pure lutetium and nitrogen at 1600\,°C \cite{Suehiro2004}. Thus, since we have not laser-heated our sample, we do not expect the formation of this compound. The EDS and WDS also support the idea that RS-LuN did not form (see Sec. S4 of the SM) since this would result in a clear signature from nitrogen as this compound is stable at ambient pressure. On the other hand, the $F\overline{4}3m$ ZB-LuN isomorph has only been predicted to form at pressures above 260\,GPa \cite{Singh2015,Chouhan2011}. Experimentally, the RS-LuN structure was shown to form preferentially when synthesised at 30\,GPa and 2000\,K \cite{Niwa2009}; that is to say, in far more extreme conditions than were attained here and in other papers, the ZB-LuN structure could not be formed, and so we do not consider it viable from hereon.

Since we do not observe any signatures of trigonal LuH$_3$ and we do not expect cubic LuH$_3$ at 2~GPa based on its predicted and observed stability \cite{Tkacz2007,Kong2012, Sun2023, Lucrezi2023, Hilleke2023}, it is likely that other lutetium hydrides have formed via the decomposition of the trigonal LuH$_3$. Firstly, hexagonal $P6_3/mmc$ LuH$_{\delta}$ compounds (0$\leq\delta\leq$0.2) form for low hydrogen concentrations \cite{Daou1965,Subramanian1982,Bonnet1976,Daou1974}. At most, these hexagonal compounds could contribute four Raman-active phonons which would help explain the low energy modes. However, our attempts to reproduce the XRD patterns with any hexagonal structure at high pressure failed. We note that, in the recovered sample at ambient pressure, we were able to identify this phase (see Sec. S3 of the SM).


The other primary lutetium hydride is $Fm\overline{3}m$ LuH$_2$, or the similar compound $Fm\overline{3}m$ LuH$_{2+x}$ with partially occupied $4b$ sites. The lattice parameter of $Fm\overline{3}m$ LuH$_2$ is reported to be $a$=5.033~\AA~at ambient conditions \cite{Yang2016,Bonnet1977,Ming2023a} which is also consistent with LuH$_{2+x}$. These phases can therefore explain the XRD pattern of the refined $Fm\overline{3}m$ phase. With regards to the Raman-activity, we expect one Raman-active $T_{2g}$ mode which was calculated to be between 960 and 1170\,\si{cm^{-1}} at ambient pressure \cite{Dangic2023}. This would be consistent with the mode measured at 1240~\icm at 1.9\,GPa. To explain our mode measured at 260\,\si{cm^{-1}}, we note that an infrared-active $T_{1u}$ mode is predicted to appear at 250\,\si{cm^{-1}} in $Fm\overline{3}m$ LuH$_3$ \cite{Dangic2023,Sun2023}. Since $Fm\overline{3}m$ LuH$_{3}$ and LuH$_{2+x}$ are structurally similar, one would expect that they share the predicted mode. LuH$_{2}$ lacks this mode \cite{Dangic2023}. Thus, provided that the $T_{1u}$ mode becomes Raman active, potentially by disorder, our excitations at 1240 and 260\,\si{cm^{-1}} could provide evidence for the presence of $Fm\overline{3}m$ LuH$_{2+x}$. Furthermore, the blue colour observed in Fig. \ref{figImage}(d) would also be consistent with the formation of $Fm\overline{3}m$ LuH$_{2+x}$, as it is also predicted to be blue \cite{Kim2023}. In summary, $Fm\overline{3}m$ LuH$_{2+x}$ is consistent with both the Raman spectra and XRD patterns we measured. However, it is clear that this phase alone cannot explain the low-energy modes since no other Raman-active modes exist, and the only other predicted $T_{1u}$ mode is at high-energy (above 1000~\icm\cite{Dangic2023,Sun2023}).

Though we identify the $Fm\overline{3}m$ structure as LuH$_{2+x}$, we still cannot explain the remaining Raman modes or the $Ia\overline{3}$ phase identified by XRD results with known phases. 
So, we shall discuss now the potential formation of the N-doped lutetium hydride compound. In Sec. S3 of the SM \cite{SI}, we show that once the pressure is released, the sample is metastable but still contains the $Fm\overline{3}m$ and $Ia\overline{3}$ phases. Most importantly, the recovered sample does not contain nitrogen as shown by both the EDS and WDS in Sec. S4 of the SM \cite{SI}.

In fact, metal nitrides are generally challenging to form due to the significant activation barrier of the non-polar, triple-bonded nitrogen atoms (bond energy 941~kJmol$^{-1}$) \cite{Luo2023}. However once synthesised, these nitrides tend to have refractory properties and are thermally and chemically stable \cite{Luo2023}. Previously, Dierkes \textit{et al.} synthesised LuN by nitriding LuH$_3$ \cite{Dierkes2017}, which is the closest analogy to the desired reaction for this work. They note that nitridation does not start below 800\,°C and even then the uptake of nitrogen is slow until above 900\,°C \cite{Dierkes2017}; they also note that LuH$_3$ begins to decompose by releasing hydrogen above 300°C. Perhaps, heating within this window under pressure would favour the formation of N-doped lutetium hydride. Cai \textit{et al.} performed a laser-heating synthesis at 1800\,°C with pure lutetium and N$_2$/H$_2$ pressure medium which formed a mixture of LuH$_2$ and LuH$_3$ with no observable nitride compounds \cite{Cai2023}. Theoretically, it has been reliably noted that there are no thermodynamically stable ternary Lu-H-N compounds: only metastable ones at best \cite{Huo2023, Sun2023, Hilleke2023,Lucrezi2023, Ferreira2023}. Furthermore, we prepared two pressure cells with pure nitrogen pressure media and we observed no change in the trigonal LuH$_3$ structure upon heating to 65\,°C at 2\,GPa followed by pressurising to 12\,GPa. This indicates that nitrogen has a limited effect on the sample; further details are provided in Secs. S2 and S3 of the SM. So based on all of this, it would seem that the synthesis, as stated in the $Nature$ paper \cite{Dasenbrock-Gammon2023}, of heating the DAC for 24\,h at 65\,°C and 2\,GPa to form N-doped lutetium hydride would be unlikely to occur.

Fortunately, with the publication of Dias' patent,
 we can gain insight into an alternative synthesis method \cite{Dias2023-patent}. According to Fig. 1 of the patent, this patentable synthesis involves heating lutetium metal in a reaction chamber with hydrogen and nitrogen gas at 4--10\,MPa and 200--400\,°C for 12--24\,h before being pressurised to 3--20\,kbar in a DAC \cite{Dias2023-patent}; this is rather different from the synthesis stated in the $Nature$ paper \cite{Dasenbrock-Gammon2023}. Despite this, our synthesis by pre-forming LuH$_3$ at 200\,°C with 4\,MPa of H$_2$ prior to loading is providentially similar, though we did not include nitrogen in this part of the synthesis. This patentable synthesis is also very similar to the work of Dierkes \textit{et al.} \cite{Dierkes2017}, though they did not heat with the two gases together in the reaction chamber at the same time. This combined with our work strongly suggests that heating the pure lutetium metal in a hydrogen and nitrogen atmosphere at high temperatures (above 200\,°C) is vital for the formation of the N-doped lutetium hydride. 
 
Overall, these considerations for the nitridation of lutetium hydride are also relevant for the partial or complete nitridation of other rare-earth hydrides and for the formation of other nitrogen compounds. Pragmatically, the successes of the rare-earth elements in producing high-temperature superconductors and the prevalence of ammonia borane syntheses have already shifted the direction of research, as evidenced by the predictions of 
nitrogen doping of rare-earth compounds \cite{Wang2020_1, Ge2021_1, Cataldo2022_1}, or simply rare-earth nitrogen compounds such as the clathrate boronitrides La(BN)$_5$ and Y(BN)$_5$ \cite{Ding2022_1}. As a result, the incorporation of nitrogen into rare-earth hydrides is a logical route of inquiry for future experimental works where the challenges of nitrogen chemistry will have to be taken into account.

In our case, we cannot conclusively say that we did or did not form N-doped LuH$_3$ at 1.9\,GPa, as it could have decomposed and ejected the nitrogen prior to the EDS and WDS measurements; however, it seems unlikely given the arguments discussed. What is clear is that at 1.9\,GPa, we formed a compound that is similar to that described by Dasenbrock-Gammon \textit{et al.} \cite{Dasenbrock-Gammon2023}, but ours was metastable and eventually decayed at ambient conditions. What is also clear is that the contradictory nature of observing many Raman-active phonons with two $Fm\overline{3}m$ lutetium lattices was an overlooked problem. Overall, the question then becomes, what is the origin of the $Ia\overline{3}$-type phase?

To explain the origin of the $Ia\overline{3}$-type phase, we speculate that this structure arises from a charge-density-wave (CDW) distortion of a pure lutetium hydride compound. Previous work on the chemically similar ScH$_3$ and YH$_3$ shows that there is an intermediate region between the ambient pressure trigonal or hexagonal structure and the high-pressure cubic phase \cite{Kume2007,Yao2010,Kume2011}. Theoretical work on YH$_3$ predicts that a Peierls distorted $C2/m$ structure forms within this intermediate phase that continues to possess a close approximation of a cubic sub-lattice \cite{Yao2010}. Unfortunately, we tried an XRD refinement of the proposed $C2/m$ structure without success, but this does not eliminate the possibility that this mechanism gives rise to other distorted structures. A similar intermediate phase was also observed in ScH$_3$ between 25 and 46\,GPa \cite{Kume2011} whereas this phase was observed in YH$_3$ between 9 and 24\,GPa \cite{Kume2007}. Since lutetium is chemically similar to scandium and yttrium, one could hypothesise that a similar intermediate Peierls-distorted/CDW phase could arise in our lutetium hydride compound. The CDW then provides a mechanism to form our $Ia\overline{3}$-type phase which is then a distortion of a higher-symmetry phase; perhaps $Fm\overline{3}m$ due to the already existing similarities. Furthermore, the pressure range of the intermediate phase seems to decrease with increasing atom size; that is to say, this intermediate phase could then coincide with our measured pressure range. It is also worth noting that a strong change in the optical gap has been observed within the CDW phase in both YH$_3$ and ScH$_3$ \cite{Kume2007, Kume2011}. As such, the observation of poor-metal behaviour and upturns in the resistivity in previous measurements on lutetium hydrides \cite{peng2023, salke2023, Ming2023a, Zhang2023} could then be evidence of a CDW phase as the gap opens. Overall, a CDW phase driving the formation of the $Ia\overline{3}$-type phase could then simultaneously explain some of the electrical properties observed, the cubic lattice of lutetium atoms, and the forest of Raman-active modes observed at low-energy without invoking the synthesis of a ternary compound. 



\section{Conclusion}

We obtain a biphasic sample which presents structural similarities to the sample of Dasenbrock-Gammon \textit{et al.} \cite{Dasenbrock-Gammon2023} by starting from pure trigonal LuH$_3$ loaded in a DAC at 1.9~GPa with a mixture of N$_2$/He. From x-ray diffraction, we clearly see a structural transformation from the initial trigonal phase to a mixture of cubic phases under pressure. Similarly, with Raman spectroscopy we observe the loss of the modes associated with the trigonal structure and see the appearance of a strong mode at 1240~\si{cm^{-1}} that we associate with the $T_{2g}$ Raman-active mode of a cubic $Fm\overline{3}m$ structure. However, we (and others) observe more excitations than are possible for two $Fm\overline{3}m$ cubic structures. Overall we believe that it is unlikely that these excitations come from impurity phases since either they are not visible in XRD, they are chemically unlikely to form, or simply their excitations do not occur in the energy range. Thus we conclude that our sample is a biphasic mixture of $Fm\overline{3}m$ LuH$_{2+x}$ and an $Ia\overline{3}$-type structure, also composed of lutetium and hydrogen, which together may describe the XRD patterns and Raman spectra. We postulate that the $Ia\overline{3}$-type structure is a distortion of a higher symmetry structure and could originate from a CDW phase. However, further theoretical work will be needed to support the origin and stability of this phase. More broadly, our discussion of nitrogen chemistry will aid future works in experimentally finding ternary compound superconductors.


\section{Acknowledgments}
This work is supported by the European Research Council (ERC) under the European Union's Horizon 2020 research and innovation program (Grant Agreement No 865826). This work has received funding from the Agence Nationale de la Recherche under the project SADAHPT. We thank Abdellali Hadj-Azzem and Elise Pachoud for lutetium preparation, and C\'{e}line Goujon for help in the preparation of the laboratory high-pressure XRD setup. We thank Laetitia Laversenne for fruitful discussions and Eva Zurek for stimulating exchanges of information.

\section*{Competing interests}
The authors declare no competing interests.

\appendix
\section*{Supplementary material}
\section{S1: Synthesis and techniques}

Lutetium (Alfa 3N) was characterised by EDS before polishing it, whereupon oxygen was clearly identified in Lu$_2$O$_3$ deposits with atomic concentrations between 20-50\,$\%$. A small amount of tantalum was also identified as shown in Fig. \ref{figEDSWDS}(a) We then polished the piece of lutetium in air until the surface became shiny instead of black in order to remove the oxide from the surface. 

\begin{figure}[h!]
\centering
\includegraphics[width=1\linewidth]{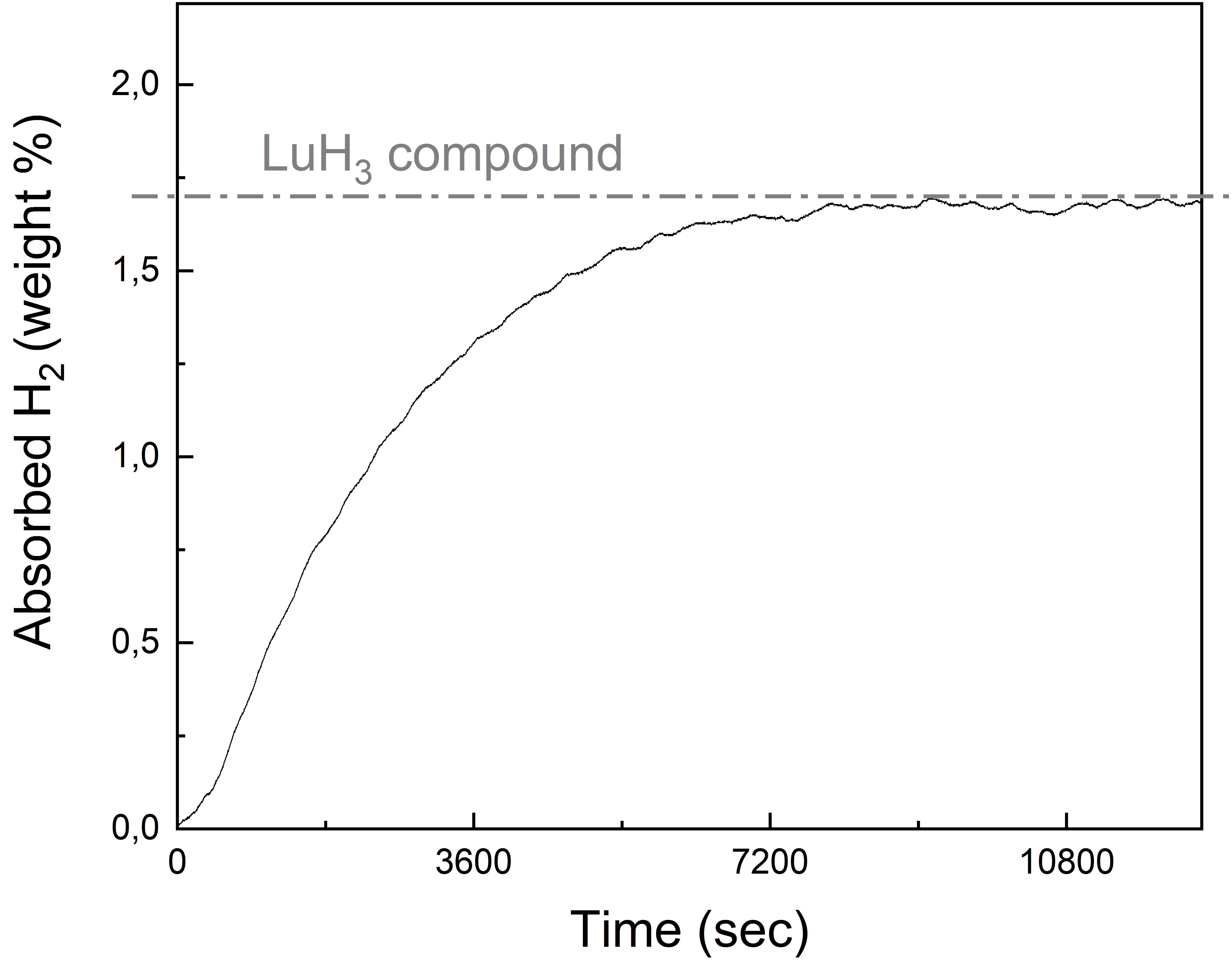}
\caption{The weight percentage of absorbed H$_2$ by lutetium as a function of time. After 3.5\,hours at 200\,$^\circ$C, 1.7\,$\%$ of absorbed H$_2$ is reached showing the successful synthesis of LuH$_3$.}
\label{synt1}
\end{figure}

LuH$_3$ was synthesised by hydrogen absorption using the Sievert method. We used a HERA C2-3000 device to measure the quantity of hydrogen absorbed (or desorbed) by the piece of lutetium as a function of time. This is calculated by measuring the hydrogen pressure variation in a sample holder of known volume. The measurement of the hydrogenation rate is performed out of equilibrium. 
The piece of polished lutetium (147.67\,mg) was placed in the sample-holder of the reaction chamber. The sample-holder and compensation chambers were then pumped for one hour at ambient temperature to remove contaminating gases. The temperature was then increased to a maximum temperature of 500\,°C at 10$^{-5}$\,mbar and kept stable for 4000\,s to outgas the container as much as possible. The temperature was then decreased to 200\,$^{\circ}$C, and H$_2$ gas at 4\,MPa was injected into the chamber. After 18\,hours, the weight percentage of absorbed H$_2$ saturated at 1.7\,$\%$ which corresponds to the expected composition of LuH$_3$, as shown in Fig.~\ref{synt1} (though only the first 3.5\,hours are shown). After the synthesis, the sample-holder was closed and transferred into an argon glove box where it was opened to recover the LuH$_3$ powder. We can qualitatively compare the hydrogen concentration within the lattice to previous measurements by comparing the $a$-axis parameter \cite{Mansmann1964, Pebler1962}. Previous work showed that a general trend amongst the trigonal/hexagonal rare-earth hydrides is that the $a$-axis parameter decreases with increasing hydrogen concentration \cite{Pebler1962}. For our sample, $a$\,=\,6.173(1)\,\AA~ whereas the $a$-axis parameter from Mansmann \textit{et al.} was determined to be 6.163\,\AA~ \cite{Mansmann1964}. Similarly, the $a$-axis value from Tkacz \textit{et al.} is 6.50\,\AA~ once converted to the equivalent hexagonal structure \cite{Tkacz2007}. Therefore, the hydrogen concentration within our sample is similar to previous samples and more densely packed than the sample of Tkacz \textit{et al.}

A thin sample of LuH$_3$ was prepared in a diamond anvil cell (DAC) with culets of 800\,$\mu$m diameter by pressing the synthesised powder between the two diamonds until the sample was approximately 5-10\,$\mu$m thick. A stainless steel gasket was indented to a thickness of 80$\,\mu$m and a hole of 400\,$\mu$m was drilled for the pressure chamber. A ruby sphere and a small piece of silicon were placed inside the pressure chamber. Prior to loading the DAC, the LuH$_3$ sample was characterised by Raman spectroscopy and X-ray diffraction (XRD) inside the unloaded DAC.

We prepared three DACs in total with the trigonal LuH$_3$ powder. The first (DAC1) was largely discussed in the main text, and we used a gas loader (Top Industrie) to load a mixture of nitrogen and helium. After purging with helium, the system was filled with 10\,bar of N$_2$ and then 1500\,bar of helium. We estimate that the quantity of N$_2$ in the pressure chamber was 4\,nmol whilst the quantity of LuH$_3$ was 11\,nmol. The DAC was then sealed at 0.1\,GPa and then we applied 1.9\,GPa and proceeded to characterise the sample by Raman spectroscopy and XRD. The second DAC (DAC2) was loaded with pure nitrogen at 1200\,bar, and the third DAC (DAC3) was cryogenically loaded with pure nitrogen at 77\,K.


 The EDS measurements used a Bruker silicon drift detector (SDD) mounted on a FESEM ZEISS Ultra+ with a working distance of 8\,mm, a take-off angle (TOA) of 35°, and an acquisition time of about 2\,mins.
To increase the chance of observing nitrogen, which emits at 0.392\,keV, WDS was performed with a JEOL-8800 Electron Probe MicroAnalyzer (EPMA/Microsonde de Castaing). Qualitative analysis of nitrogen used a LDE1H synthetic superlattice analyzing crystal (Si/W) and TAP for Lu-M$\alpha$ (1.58\,keV). On the EPMA system, the TOA is 40°. For the EDS, the electron-beam was rastered over an area of approximately 2x2\,$\mu m^2$, whilst for the WDS a defocussed spot of 10\,$\mu m$ was used to limit the material degradation by overheating or carbon contamination from the adhesive tape. Both experiments used several voltages (from 5-15\,keV) though the ionisation efficiency of nitrogen is enhanced at low voltages. 



\begin{figure}[h!]
\centering
\includegraphics[width=1\linewidth]{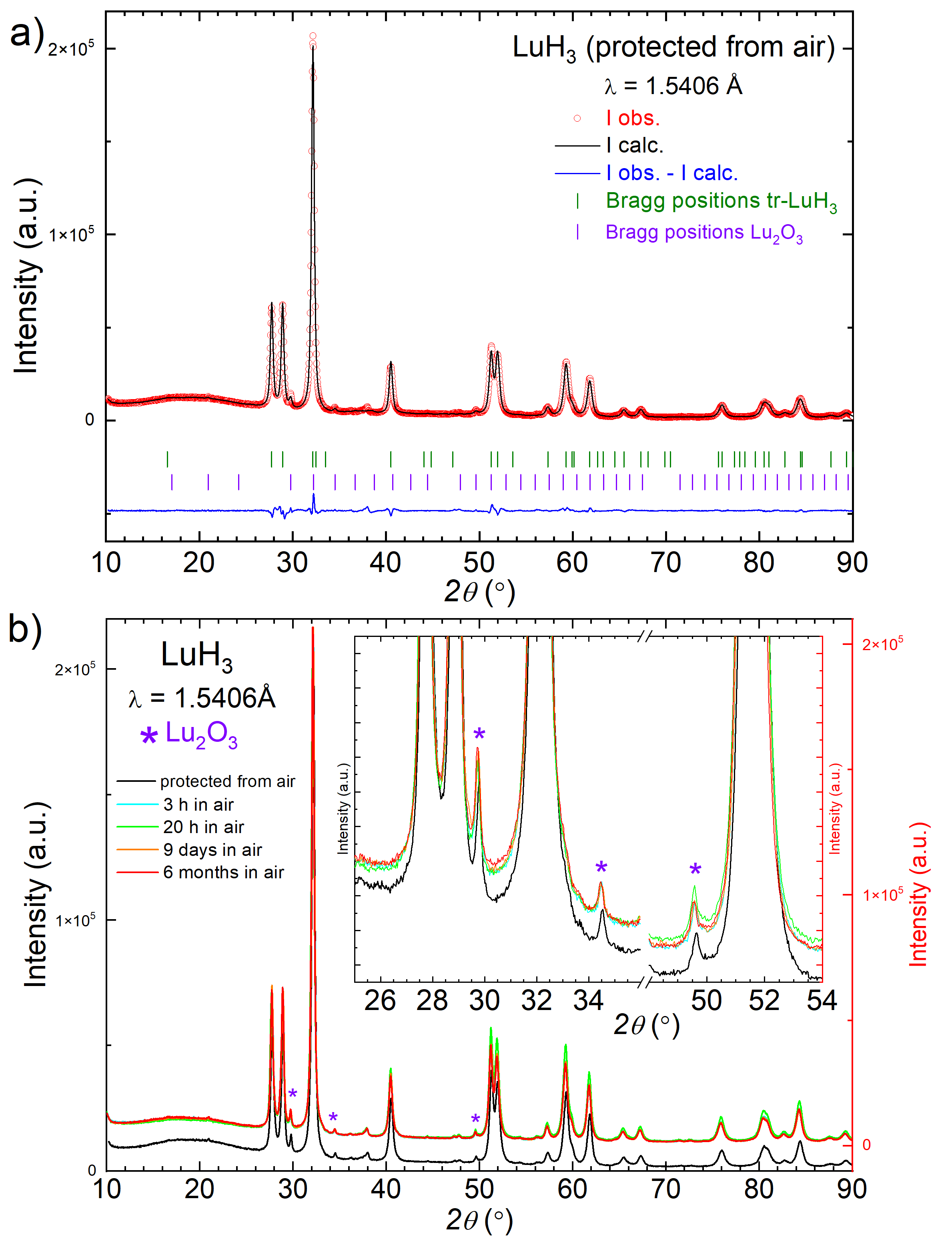}
\caption{Powder X-ray diffraction on the trigonal LuH$_3$. (a) Sample kept in glove-box and sealed between two pieces of tape during the measurement. The reliability values for the fit are (R$_p$=7.25\%, R$_{wp}$=7.95\%, R$_{exp}$=1.79\%, $\chi^2$=19.7). (b) Even after 6 months of exposure to air, the quantity of Lu$_2$O$_3$ impurities did not change significantly with time. The patterns are shifted by a constant offset to aid comparison.}
\label{figXRD1}
\end{figure}
\vspace{6mm}

X-ray powder diffraction of the starting LuH$_3$ was performed immediately after the hydrogenation of lutetium using a D5000T diffractometer (Cu-K$\alpha$ radiation), at ambient pressure (and outside the DAC). The measurement was repeated several times (up to 9\,days after the first measurement, and a final time after 6\,months) to determine the effect of air exposure on LuH$_3$. The Rietveld refinements were done with FullProf software \cite{FullProf}. The X-ray powder diffraction after loading at 1.9\,GPa in DAC was performed on the ESRF beamline ID15B with $\lambda$\,=\,0.411\,\AA. Additional measurements on the LuH$_3$ powder at ambient pressure were also performed on the same ESRF beamline. Calibration of the detector-to-sample distance, beam orientation, detector tilt with respect to the omega rotation axis, and the used wavelength was determined by a Si powder standard (‘NIST 640 C’ from NIST). The X-ray beam was focused to 4x3\,$\mu$m$^2$ using Be compound refractive lenses. 2D images were collected with a six degrees oscillation of the DAC using an Eiger 2X CdTe 9M photon counting detector from Dectris and integrated into a 1D pattern using the Dioptas software \cite{Prescher2015}. Le Bail refinements (lattice parameter, peak profile, and background) on the loaded DAC at 1.9\,GPa were done using the GSAS-2 package \cite{Toby2013}.

Polarised Raman scattering was performed in quasi-backscattering geometry at 300\,K with an incident laser-line at 532\,nm from a solid-state laser. The DAC was placed in a vacuum to avoid measuring the Raman response of air. We used a laser power between 2.5-10\,mW with a typical spot size of 25\,$\mu$m. The scattered light was analysed by a single grating and a triple grating subtractive spectrometer, both were equipped with liquid nitrogen-cooled CCD detectors. The crossed and parallel polarisation dependence was measured by changing the orientation of the polariser on the collection path. We measured the Raman signal of pure LuH$_3$ in the DAC before and after loading the pressure medium. 

\section{S2: Trigonal lutetium trihydride}

Fig. \ref{figXRD1}(a) shows the pattern of the lutetium trihydride immediately after synthesis; it is well-described by a trigonal structure with some Lu$_2$O$_3$ impurities. After the first XRD measurement, we left a small fraction of the LuH$_3$ powder exposed to air and measured the XRD several times over the course of 9\,days to check its stability. The rest of the powder was immediately stored under vacuum or in an argon glove box. Figure \ref{figXRD1}(b) shows that despite being in contact with air, the Lu$_2$O$_3$ content is similar within the error bar, i.e. 3.4(1)\,$\%$ vs 3.2(1)\,$\%$ from before. This remains true after 6\,months of exposure to air.

\begin{figure}[h!]
\centering
\includegraphics[width=1.05\linewidth]{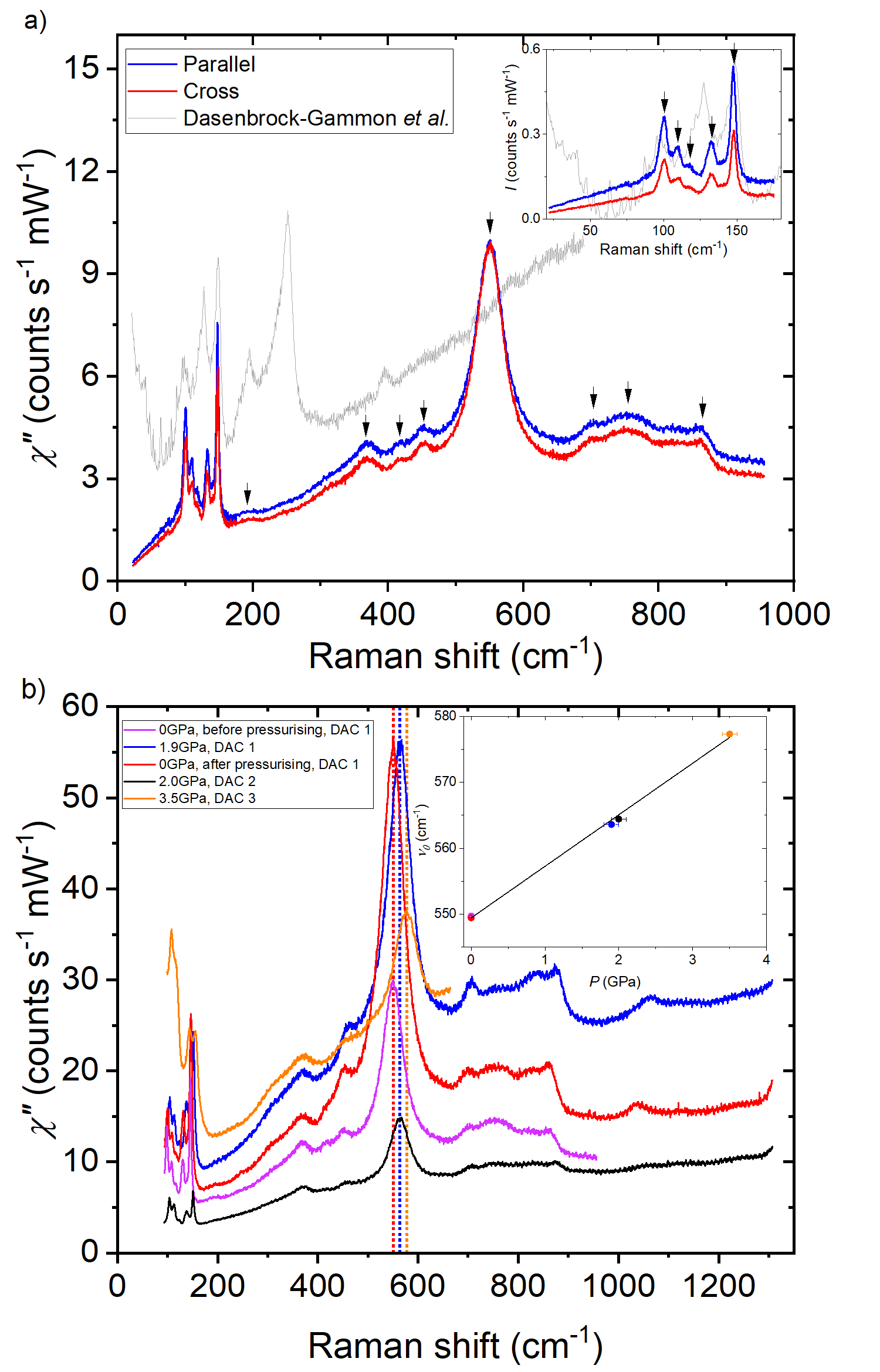}
\caption{(a) Raman susceptibility of trigonal LuH$_3$ at 300\,K and 1\,bar measured in the unloaded DAC1 in cross and parallel configurations. Arrows point to features of interest. Below 175\,\si{cm^{-1}}, data are scaled to overlay with the single-stage data measured at higher energies. The inset shows the unscaled triple-stage data at low energy. The raw data at ambient pressure from ref \cite{Dasenbrock-Gammon2023} are shown in grey and are scaled to aid comparison. (b) The pressure evolution of the translucent part of the sample at 300\,K in DAC1. The translucent part remained trigonal throughout the pressure cycle: from 0 to 1.9\,GPa and back to 0\,GPa. Scaled comparisons with two other samples in DAC2 and DAC3 (nitrogen pressure medium) at 2.0\,GPa and 3.5\,GPa respectively are shown. Dotted lines show the Raman shift of the dominant peak at ambient (red) and high pressure (blue and orange). Inset shows the pressure dependence of the dominant peak and a linear fit.}
\label{figRaman0}
\end{figure}

Fig \ref{figRaman0}.a shows the polarisation dependent Raman spectra of the ambient pressure trigonal LuH$_3$ below 955\,\si{cm^{-1}}; at higher energies we do not identify any excitations that clearly originate from LuH$_3$. Within the aforementioned range, we observe 13 features (marked by arrows) which could account for most of the expected 17 phonons of the $P\overline{3}c1$ trigonal structure. 
Overall, we do not observe any significant differences between the different polarisations. The inset shows the low-energy spectra down to 20\,\si{cm^{-1}} where we do not see any more notable features.

Fig. \ref{figRaman0}(b) shows the Raman spectrum of trigonal LuH$_3$ before pressurising alongside the spectra of the translucent part of the sample at high pressure and again at ambient pressure after releasing the pressure. Apart from a hardening of the phonon modes under pressure, we do not see any drastic change in the spectra. Importantly, the number of modes observed does not change over the pressure cycle, so it seems that this part of the sample was untransformed and largely unaffected by the high pressure. Why remains unclear.

\begin{figure}[h!]
\centering
\includegraphics[width=0.9\linewidth]{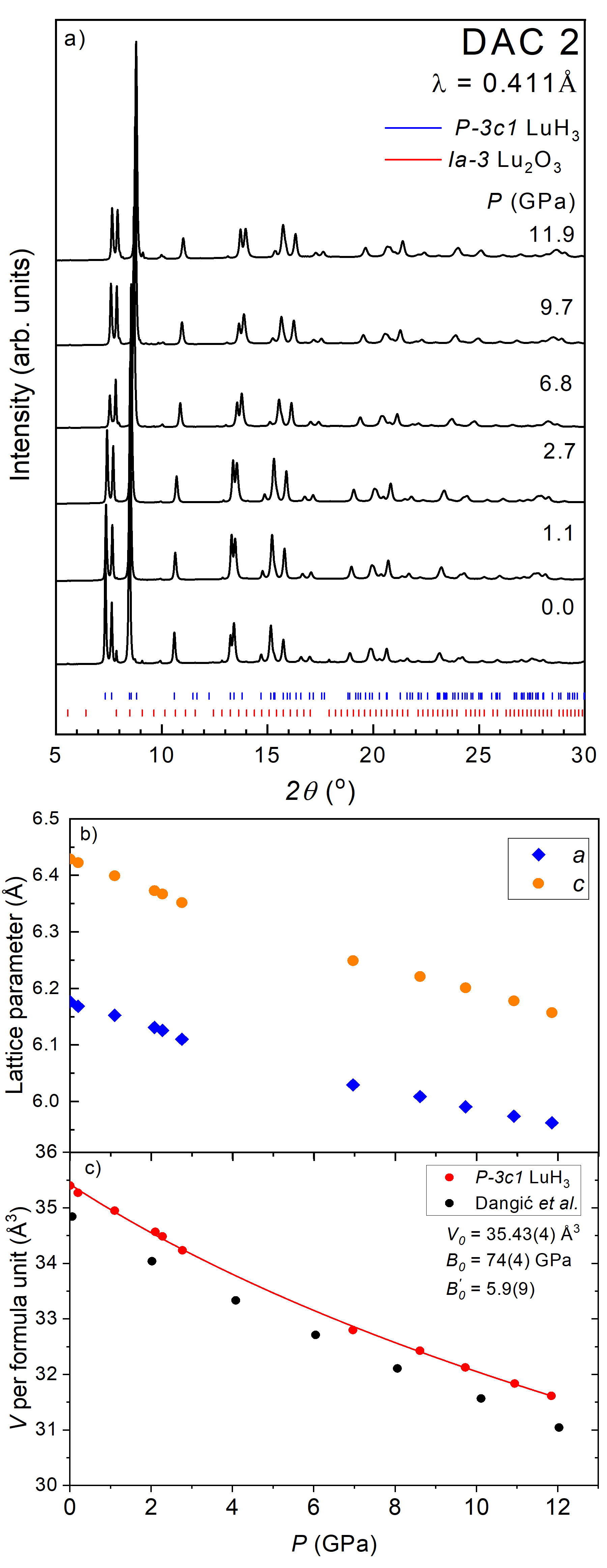}
\caption{(a) X-ray diffraction patterns of the trigonal LuH$_3$ phase under pressure in DAC2. (b) Variation of the fitted lattice parameters with pressure for the trigonal phase. (c) The lattice volume versus pressure data for the trigonal phase with a Birch-Murnaghan fit. Predictions by Dangi\'c \textit{et al.} \cite{Dangic2023} are also shown for comparison.}
\label{figVvsP}
\end{figure}

 DAC2 was primarily used to determine the pressure dependence of the lattice volume. Initially, this was followed at the ESRF Beamline up to 3\,GPa. Since the sample remained trigonal, it was heated to 65\,°C and during this process, the pressure increased up to 7\,GPa, yet the sample remained trigonal. The pressure was then increased further until 12\,GPa at room temperature with XRD being measured at every pressure and the result is shown in Fig. \ref{figVvsP}(a). The lattice parameters of the refined trigonal structure are shown in Fig. \ref{figVvsP}(b), whilst Fig. \ref{figVvsP}(c) shows the volume dependence on pressure. This was also calculated by Dangi\'c \textit{et al.} which is presented alongside the volume dependence determined here and shows a similar trend with a small offset. After that, the pressure was decreased to 2\,GPa, whereupon the Raman spectroscopy was measured which is presented in figure \ref{figRaman0}(b) Throughout all of the pressure changes the sample remained trigonal. 

\begin{figure}[h!]
\centering
\includegraphics[width=1\linewidth]{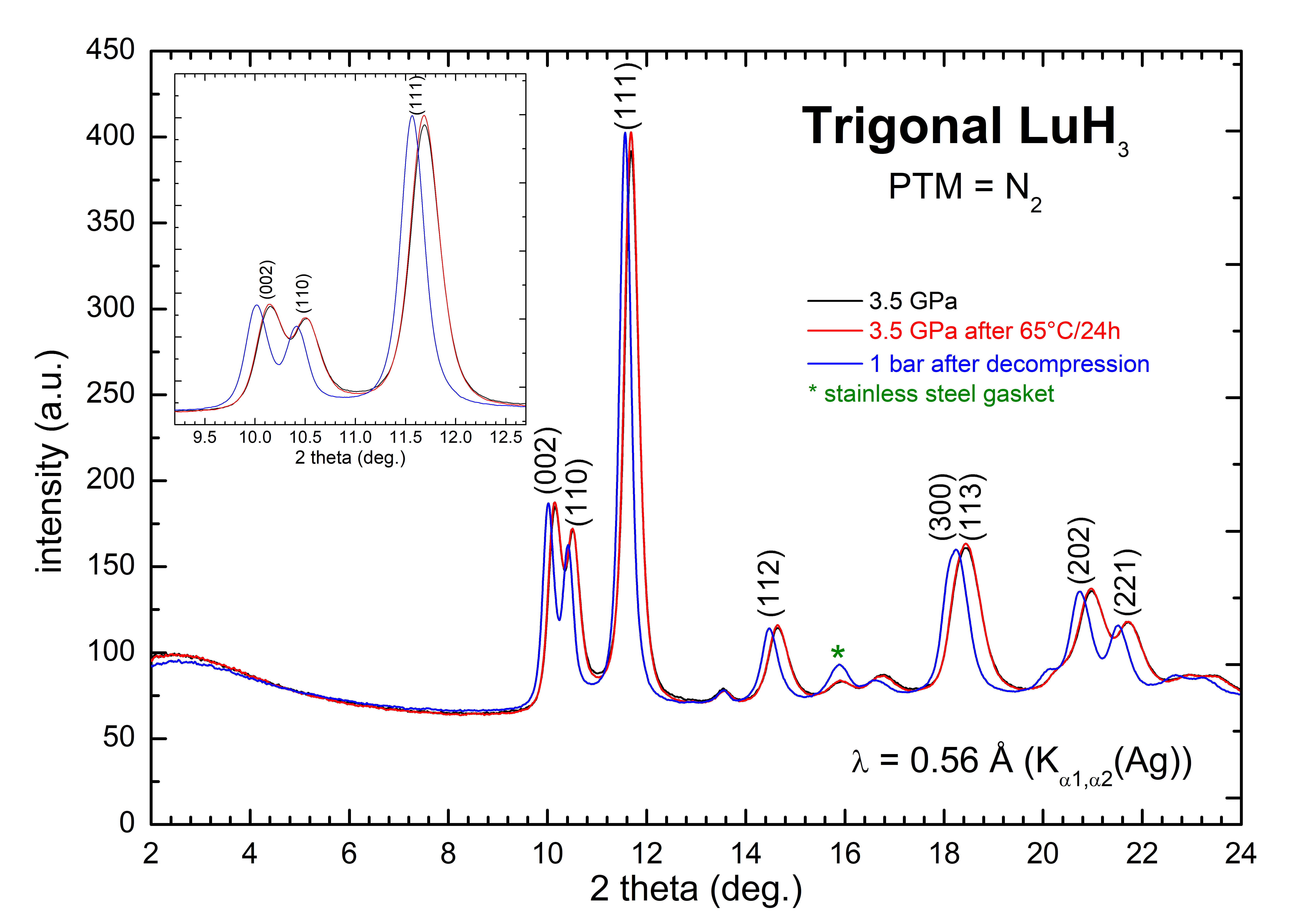}
\caption{Powder X-ray diffraction of the trigonal LuH$_3$ in cryogenically loaded nitrogen pressure medium (DAC3). Black and red lines show the data at 3.5\,GPa before and after heating respectively; they are effectively identical and overlay almost perfectly. Blue data show the pattern after releasing the pressure.}
\label{figXRDCryogenic}
\end{figure}

After cryogenically loading DAC3 and warming to room temperature, the pressure was determined to be 3.5\,GPa. At this pressure, both the Raman and XRD confirmed that the structure remained trigonal (see Figs. \ref{figRaman0}(b) and \ref{figXRDCryogenic} respectively). Here a laboratory K$\alpha$-Ag source ($\lambda$\,=\,0.56\,\AA) was also used to measure the XRD. The DAC was then heated at 65\,°C for 24\,h as was done for both the sample in the main text and the Dasenbrock-Gammon sample \cite{Dasenbrock-Gammon2023}; the resulting XRD pattern is shown in Fig. \ref{figXRDCryogenic} and there is no measurable difference within the error, as shown by the refined lattice parameters in table \ref{TabXRDCryoN2}. Overall we do not reproduce the cubic structural transition in this cell either. Upon decompression, the recovered sample remained trigonal but with a slightly larger $a$-axis than the original sample before compression, though this could be intrinsic hysteretic behaviour of the sample caused by compression and decompression.

\begin{table}[]
\begin{tabular}{c|c|c|c } \hline \hline
Conditions          & $\lambda$ (\AA) & $a$-axis (\AA)       & $c$-axis (\AA)    \\ \hline
Before (1bar/300K)  & 1.54              & 6.1680(8)               & 6.422(1)             \\ 
3.5\,GPa before heating     & 0.56              & 6.111(5)                & 6.335(9)             \\ 
3.5\,GPa after heating & 0.56              & 6.113(6)                & 6.338(9)             \\ 
After decompression & 0.56              & 6.1744(4) & 6.421(8) \\ \hline \hline
\end{tabular}
\caption{Refined attice parameters of the LuH$_3$ sample loaded with cryogenic liquid nitrogen (DAC3) at several stages throughout the synthesis process.}
\label{TabXRDCryoN2}
\end{table}

In both cells loaded with pure nitrogen (DAC2 and DAC3), we observe Raman spectra that resemble trigonal LuH$_3$ at high pressure, as shown by Fig. \ref{figRaman0}. These trigonal samples and the small trigonal part in DAC1 all show a very similar linear hardening with pressure for the dominant phonon mode, as shown by the inset of Fig. \ref{figRaman0}(b). Sato \textit{et al.} showed that the Raman spectra of pressurised SiO$_2$ glass change when in a helium pressure medium, as the helium atoms occupy interstitials within the silicate network \cite{Sato2011a}. Here, we do not observe any significant difference between the trigonal LuH$_3$ samples loaded in the pressure media, and the hardening of the phonons under pressure follows the same behaviour in all of the pressure media. This leads us to believe that the helium pressure medium is not causing the structural change in DAC1. Considering the effects of the pressure media themselves, since both helium and nitrogen perform well as hydrostatic pressure media to at least 10\,GPa \cite{Klotz2009}, we do not expect significant uniaxial effects below 2\,GPa. So the difference in hydrostaticity is unlikely to be the origin of the difference between DAC1 (with transformation) and, DAC2 and DAC3 (without transformation).

\section{S3: Transformation of the LuH$_3$ sample}

\begin{figure}[]
\centering
\includegraphics[width=0.98\linewidth]{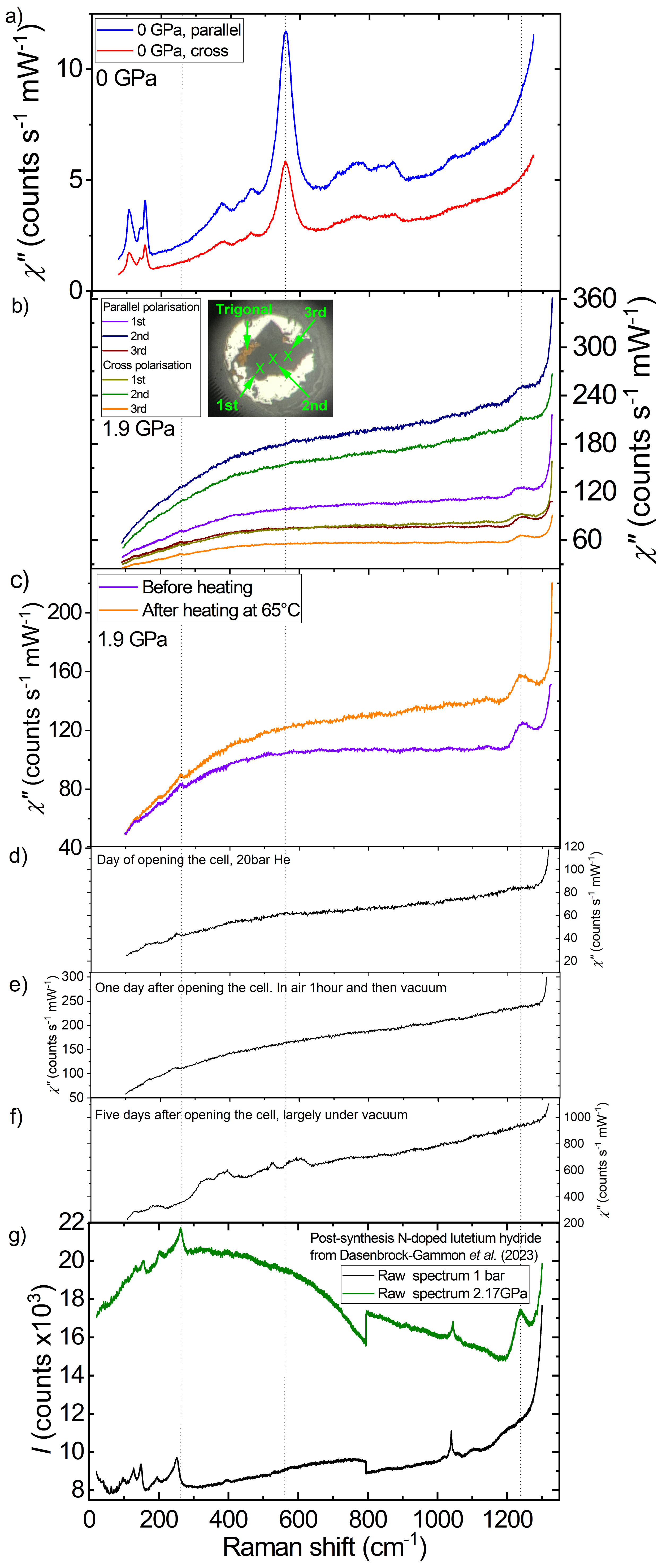}
\caption{(a) Raman susceptibility of the trigonal LuH$_3$ at ambient pressure. (b) Raman susceptibility of the compound at 1.9\,GPa (DAC1). Data on three different spots are presented, and the inset shows the locations of the spots on the sample. Below 175\,\si{cm^{-1}}, triple-stage data are overlaid on the high-energy spectra. (c) Raman susceptibility of the high-pressure phase before and after annealing at 1.9\,GPa and 65\,$^{\circ}$C for 24\,h. The purple data were scaled such that the intensity of the high-energy mode is similar. (d) through to (f) show the Raman susceptibility of the annealed sample at several times after opening the DAC. (g) The raw Raman spectra of part A of the sample from Dasenbrock-Gammon \textit{et al.} at ambient pressure and at 2.17\,GPa are presented \cite{Dasenbrock-Gammon2023}.}
\label{figRaman2SI}
\end{figure}

Figs. \ref{figRaman2SI}(a) and \ref{figRaman2SI}(b) show wide-range Raman spectra on the ambient pressure trigonal LuH$_3$ and the high-pressure compound of DAC1. Here the modes in the high-pressure structure clearly do not resemble the modes in the trigonal structure. The spectra of the high-pressure phase for multiple spots on the sample also show the same features, though the background does change. The locations of these different spots are shown in the inset image. In table \ref{TabEnergies}, we write the energies of the excitations seen in the original trigonal LuH$_3$ and the high-pressure compound.

\begin{table}[h!]
\footnotesize 
\begin{tabular}{c|c|c}
\hline \hline
Compounds & LuH$_3$       & High pressure compound  \\
& (0~GPa)&(1.9~GPa) \\
\hline
 Energy& 100.2 & 128         \\ 
 (cm$^{-1}$)& 109.4&  164    \\ 
     & 117.4&    202  \\ 
     & 132.6&    260 \\ 
     & 147.5&    \textit{1141} \\ 
    & 368.4&     1241     \\ 
     & 416.8&          \\ 
     & 454.2&          \\ 
     & 550.2&          \\ 
    &702.2 &          \\ 
    & 755&          \\ 
    &\textit{829}&          \\ 
    & 861.8 &          \\ 
    & \textit{1039}&            \\    
\hline \hline
\end{tabular}
\caption{Raman modes energy measured on trigonal LuH$_3$ at ambient pressure and the high-pressure compound measured at 1.9\,GPa in DAC1. In italics, are the modes which are difficult to identify. }
\label{TabEnergies}
\end{table}



To complete the synthesis as described by Dasenbrock-Gammon \textit{et al}, DAC1 was heated at 65\,$^{\circ}$C for 24\,h at 1.9\,GPa. Fig. \ref{figRaman2SI}(c) shows the resulting Raman spectra; not much has changed. 

\begin{figure}[]
\centering
\includegraphics[width=1\linewidth]{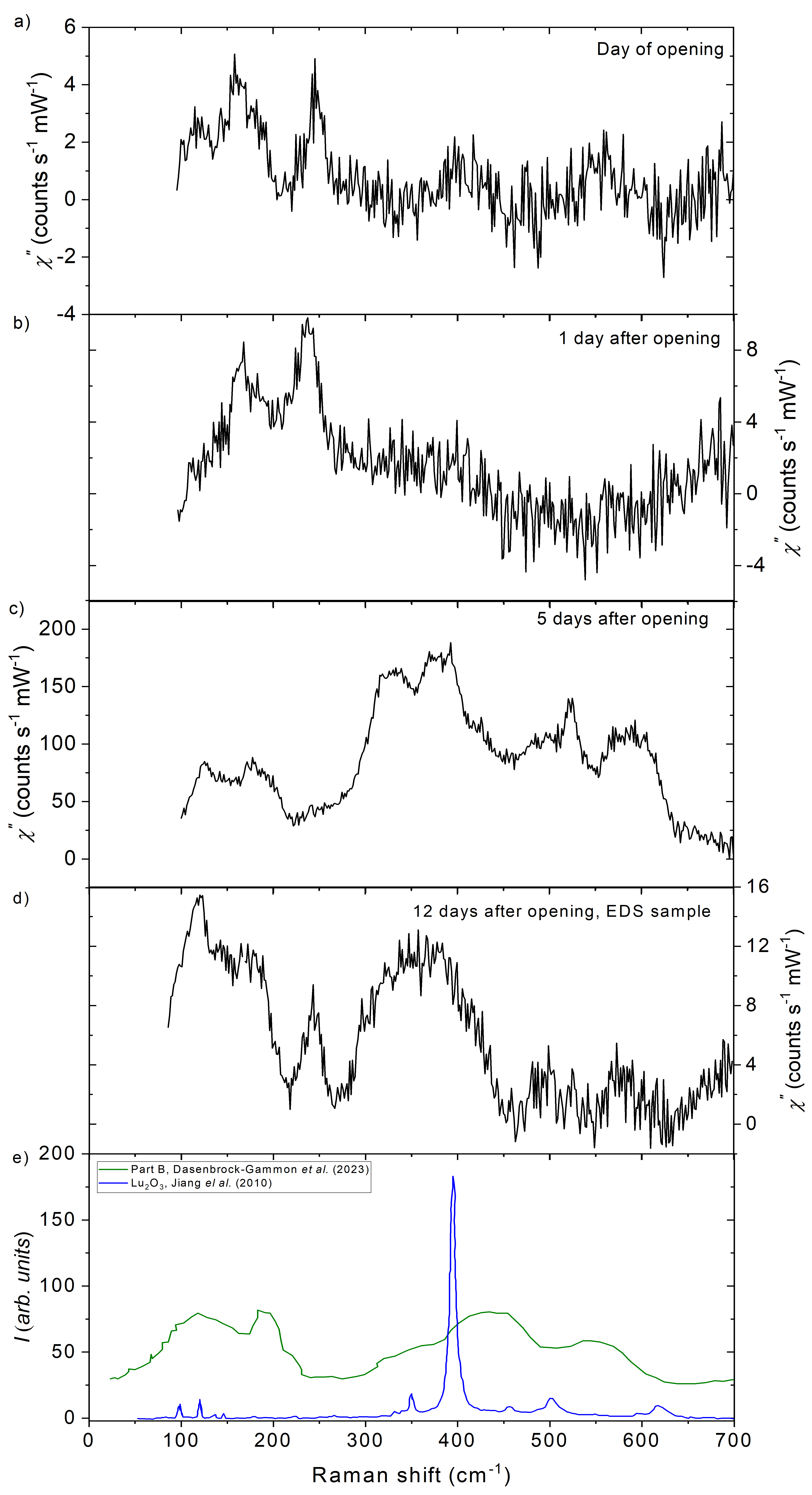}
\caption{Polynomial background subtracted Raman spectra showing the time evolution of the sample at 0\,GPa after pressurising to 1.9\,GPa and heating to 65\,°C. We also overlay data from Dasenbrock-Gammon \textit{et al.} \cite{Dasenbrock-Gammon2023} and Jiang \textit{et al.} \cite{Jiang2010a} to compare against `part B' of their sample and Lu$_2$O$_3$, respectively.}
\label{RamanSI_FinalPhaseDecomp}
\end{figure}

\begin{table}[]
\footnotesize
\begin{tabular}{c|c|c|cc}
\hline \hline
\multicolumn{1}{c|}{} & $Fm\overline{3}m$-type  & $Ia\bar{3}$-type  & \multicolumn{2}{c}{$P6_3/mmc$ LuH$_{x}$}            \\ \hline
                       & $a$-axis (\AA) & $a$-axis (\AA) & \multicolumn{1}{c|}{$a$-axis (\AA)} & $c$-axis (\AA) \\ \hline
1                      & 4.798       & 10.427     & \multicolumn{1}{c|}{3.529}       & 5.588       \\ \hline
2                      & 4.806       & 10.433     & \multicolumn{1}{c|}{-}           & -           \\ \hline
3                      & 4.776       & -          & \multicolumn{1}{c|}{3.515}       & 5.589       \\ \hline
4                      & 4.773       & -          & \multicolumn{1}{c|}{3.5099}      & 5.584       \\ \hline
5                      & 4.796       & 10.402     & \multicolumn{1}{c|}{-}           & -           \\ \hline
6                      & 4.785       & 10.409     & \multicolumn{1}{c|}{3.527}       & 5.561       \\ \hline
7                      & 4.781       & 10.399     & \multicolumn{1}{c|}{-}           & -           \\ \hline
8                      & 4.788       & 10.410     & \multicolumn{1}{c|}{3.524}       & 5.583       \\ \hline
Average                & 4.79(1)     & 10.41(1)   & \multicolumn{1}{c|}{3.521(7)}    & 5.58(1)     \\ \hline\hline
\end{tabular}
\caption{Lattice parameters from Le Bail refinements of the three phases in the sample from DAC1 released 
 at ambient pressure and measured in several different locations on the sample. A hyphen means that the given phase was not observed in that location.}
\label{TabReleasedSample}
\end{table}

We then opened the DAC1 in helium gas (20\,bar) to avoid contact with air. Then we slightly closed the DAC to keep the sample in a helium environment and remeasured the sample at essentially ambient pressure. The results are shown in Figs. ~\ref{figRaman2SI}(d) to \ref{figRaman2SI}(f). Shortly after opening, the spectrum resembles the cubic phase with a peak located just below 250\,\si{cm^{-1}} and what could be a broad and weak remainder of the peak at 1240\,\si{cm^{-1}}. However after one day, this high-energy peak has disappeared but the low-energy peak remains. Fig. \ref{figRaman2SI}(f) shows the spectrum after several days (during which time the sample was stored under vacuum), and clearly the structure has changed once again. This spectrum resembles neither the cubic nor the trigonal phase. In Fig. \ref{RamanSI_FinalPhaseDecomp}, we compare the background-subtracted signals of the data in Figs. \ref{figRaman2SI}(d)-(f) against the spectra of Lu$_2$O$_3$ \cite{Jiang2010a} and `part B' from Dasenbrock-Gammon \textit{et al.} \cite{Dasenbrock-Gammon2023}. There is no strong resemblance between either of the other compounds, with the exception of the most intense peak of Lu$_2$O$_3$, which would have to be considerably broadened, and the low-energy peaks of `part B', but the rest of the spectrum is different.


\begin{figure}[h!]
\centering
\includegraphics[width=0.8\linewidth]{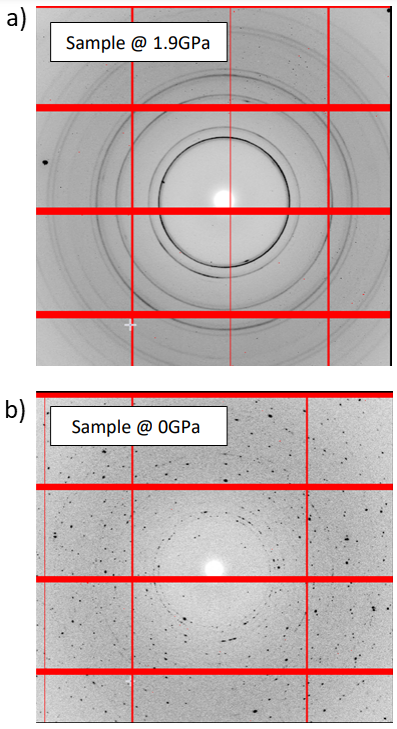}
\caption{2D XRD images of the sample after heating to 65\,°C (a) at 300\,K and 1.9\,GPa and (b) after the pressure was released. Temporally, (b) was measured between the Raman data displayed in Figs. \ref{RamanSI_FinalPhaseDecomp}(e) and \ref{RamanSI_FinalPhaseDecomp}(f), i.e. between 1\,day and 5\,days after opening the cell. Both XRD images were obtained with the sample in the DAC.}
\label{XRD4}
\end{figure}

\begin{figure}[h!]
\centering
\includegraphics[width=1\linewidth]{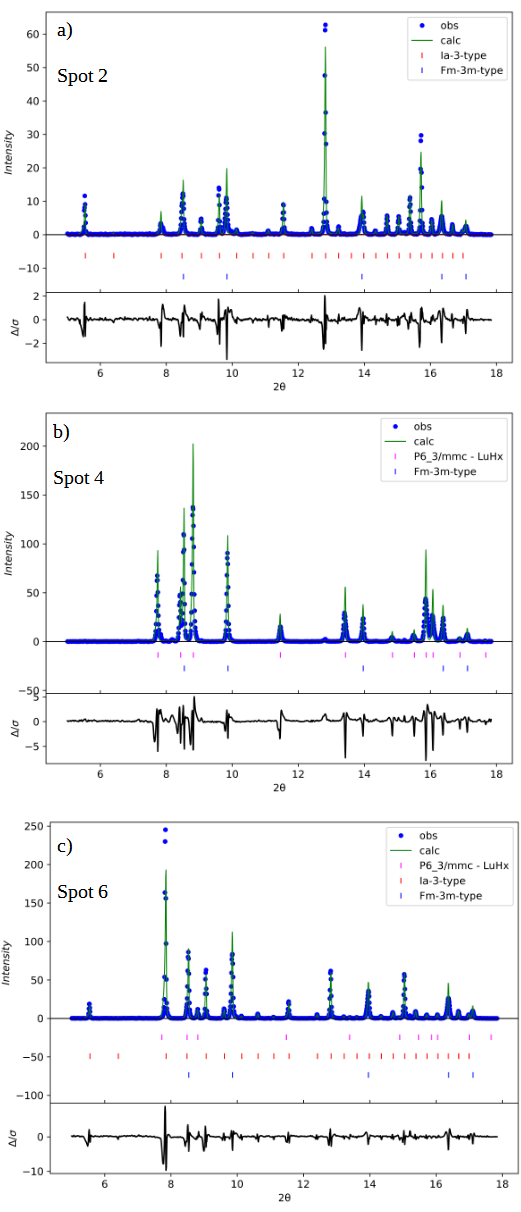}
\caption{XRD patterns on three different spots of the released sample that were measured at the ESRF with $\lambda$\,=\,0.411\,\AA. We identify three different phases: $Fm\overline{3}m$-type, $Ia\bar{3}$-type, and $P6_3/mmc$ LuH$_x$ which are all fitted with Le Bail fits.}
\label{XRDReleasedSample}
\end{figure}

Subsequently, we measured the XRD after releasing the pressure, and the corresponding diffraction XRD image is shown in Fig. \ref{XRD4}(b), whereas Fig. \ref{XRD4}(a) shows the high-pressure phase. The most evident change is that the 0\,GPa XRD image has become spotty instead of forming continuous rings. This shows that the crystalline domain sizes are larger than the X-ray beam size (4x3\,µm$^2$) which means that we can no longer fit the patterns with Rietveld refinements. Qualitatively, in the ambient pressure patterns, we see three phases as shown in Fig. \ref{XRDReleasedSample}. We measured 8 different spots. Firstly, we observe similar $Fm\overline{3}m$ and $Ia\bar{3}$-type structures to those measured at high pressure, but in addition we observe a $P6_3/mmc$ phase. $Fm\overline{3}m$ phase is present in every measured spot, but this forms either a biphasic mixture with the $Ia\bar{3}$-type (3/8 spots) or the hexagonal phase (2/8 spots), or it forms a triphasic mixture (3/8 spots). The refined lattice parameters of the measured phases in different locations are shown in table \ref{TabReleasedSample}.

To understand this, we must first consider the binary mixture phase diagram of lutetium and hydrogen \cite{Daou1965, Subramanian1982,Bonnet1971,Daou1974}. For low hydrogen concentrations up to 0.2\,H/Lu, a pure hexagonal $P6_3/mmc$ LuH$_{\delta}$ ($0\leq\delta\leq0.2$) forms; the lattice parameters of which increase with increasing hydrogen concentration until they saturate at $a$=3.5240\,\AA~and $c$=5.6050\,\AA~ for LuH$_{0.2}$ \cite{Bonnet1971}. Both of our average values of $a$=3.521(7)\,\AA~and $c$=5.58(1)\,\AA~indicate a lower hydrogen concentration: the values of $a$ and $c$ imply $\delta$=0.16(7) and $\delta$=0.09(3), respectively. Beyond 0.2\,H/Lu, a binary mixture of the $P6_3/mmc$ LuH$_{\delta}$ and an $Fm\overline{3}m$ phase forms. There is uncertainty where the end of this binary mixture ends: some sources say $\approx0.6$\,H/Lu \cite{Daou1965, Subramanian1982} while another says 1.8\,H/Lu \cite{Daou1974}. The latter concentration forms a compound that is approximately the same as LuH$_2$ which has a lattice parameter of $a$=5.035~\AA~ \cite{Bonnet1971}. This value is much larger than our average value of 4.79(1)\,\AA. But in the instance that $0.6$\,H/Lu is the beginning of the binary mixture, it is then probable that the low concentration $Fm\overline{3}m$ phase would have a much smaller lattice parameter than LuH$_2$ which could then be close to our value. Finally and as discussed in the main text, the lattice parameter of the $Ia\bar{3}$-type structure expands when the pressure is released and becomes larger than the ambient pressure value of 10.38\,\AA\, for Lu$_2$O$_3$, therefore we conclude that the $Ia\bar{3}$-type phase is a distinct compound from Lu$_2$O$_3$.

Here and in the main text, we consider the decomposition of the initial LuH$_3$ into lower hydrides. This must result in the formation of H$_2$ which should in principle be measurable by Raman spectroscopy. At high energy, there exists a well-known hydrogen vibron excitation at approximately 4200\,\si{cm^{-1}} at low pressure and 300\,K \cite{Eremets2011_1,Eremets2023_1, Sharma1980}. However, this vibron is inherently weak and generally only visible with a pure hydrogen pressure medium or with ammonia borane after laser heating due to the large concentration of hydrogen present. In our work, the proposed decomposition of LuH$_3$ to LuH$_{2+x}$ would only produce a fraction of a hydrogen atom per unit cell and therefore a low concentration of hydrogen; thus the intensity of the vibron will be weaker. Furthermore, the hydrogen can escape the pressure cell which further reduces the quantity present and diminishes the intensity. As a result of all of these reasons, we did not observe the high-energy hydrogen vibron. There also exists a weaker hydrogen excitation at approximately 1044\,\si{cm^{-1}} \cite{Sharma1980}, which is clearly observable in the data of Dasenbrock-Gammon \textit{et al.} in Fig. 3(c) of the main text. This is due to their use of a hydrogen pressure medium, but despite that, the excitation remains weak. Since we did not use a hydrogen pressure medium and the aforementioned reasons, it is not surprising that we do not observe it.



\subsection{S4: EDS and WDS analysis of the recovered sample}

\begin{figure}[h!]
\centering
\includegraphics[width=1\linewidth]{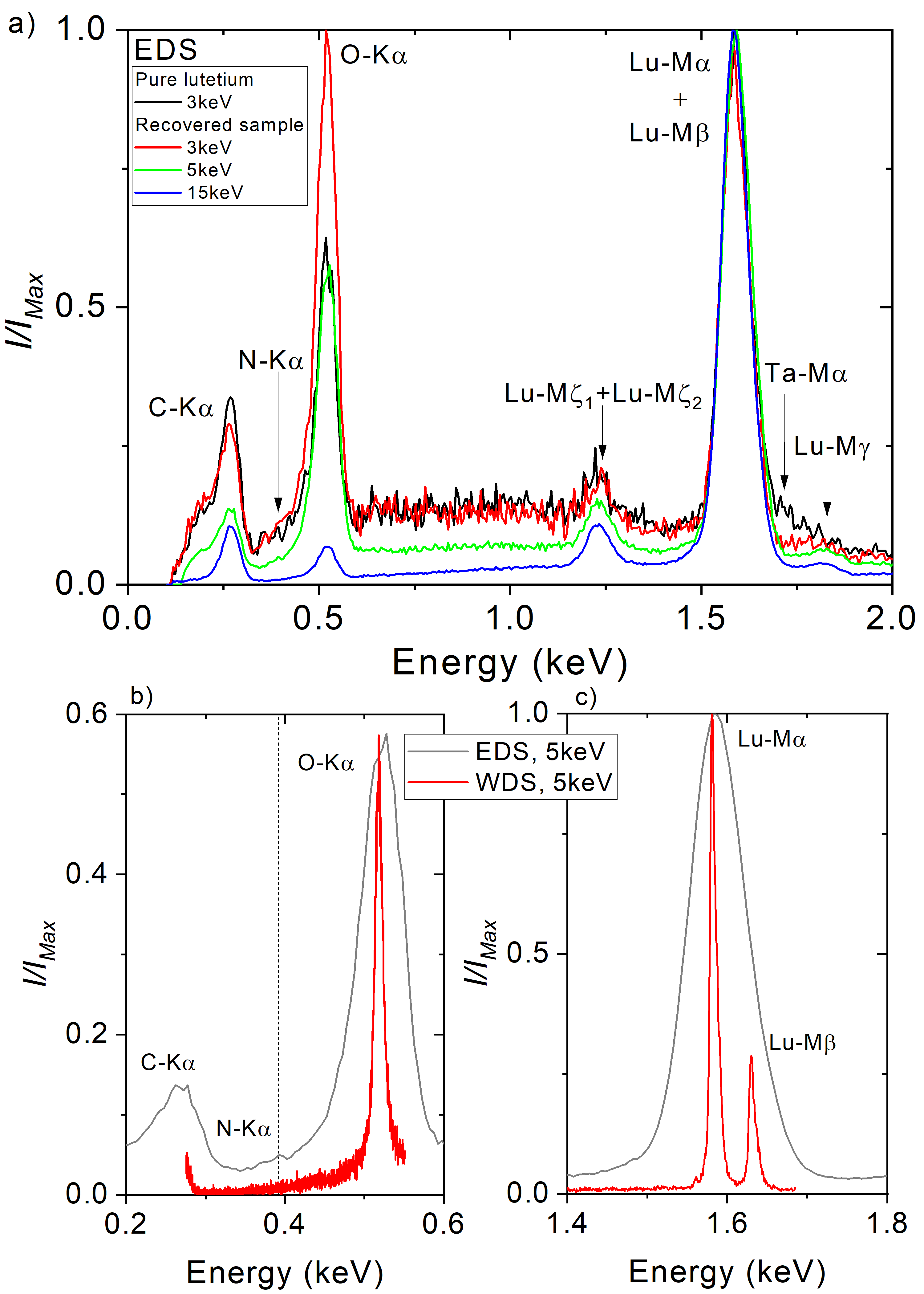}
\caption{(a) EDS measurements on the high-pressure sample of DAC1 after releasing the pressure and polished pure lutetium at different accelerating voltages. Several key emission lines are indicated. (b) and (c) A comparison between the 5\,keV measurements with EDS and WDS on the recovered sample. The WDS measurements were scaled on the O-K$\alpha$ and Lu-M$\alpha$ lines respectively to aid comparison.}
\label{figEDSWDS}
\end{figure}

Scanning electron microscopy with X-ray energy dispersive spectroscopy (EDS) and wavelength dispersive spectroscopy (WDS) were used to analyse the composition of the pure lutetium and recovered sample. Fig. \ref{figEDSWDS}(a) shows the EDS spectra of the recovered sample after pressuring at 1.9\,GPa and heating at 65\,°C, and pure Lu after polishing; all spectra were normalised by the maximum intensity of a given spectrum. At high accelerating voltages, one preferentially excites the energy levels of the heavier atoms, whilst at low voltages, the signal of lighter elements becomes more intense. This is most clearly seen in the intensity of the O-K$\alpha$ peak which grows in intensity relative to the Lu-M$\alpha$ peak at low voltages. Thus to find nitrogen, lower accelerating voltages should be used.

Firstly though we should comment on the other atoms detected: oxygen and carbon. As mentioned before, oxygen originates from Lu$_2$O$_3$ and is also present in freshly polished lutetium metal. Its presence is not a surprise. The carbon originates from the tape used to attach the sample to the sample holder, as a conductive surface must be used, therefore this is also expected.

The characteristic K$\alpha$ emission energy of nitrogen is situated at 0.392\,keV as indicated in fig \ref{figEDSWDS}.a. However, within the noise of the measurement for these EDS measurements, there is no indication of nitrogen in the structure. We also note that there is very little difference between the recovered sample and the pure lutetium. We also used WDS which has superior resolving power, as shown in Figs. \ref{figEDSWDS}(b) and  \ref{figEDSWDS}(c) by the narrower O-K$\alpha$ line and the Lu-M$\alpha$+Lu-M$\beta$ line being clearly resolved into the two spectral lines. This helps to distinguish the potential nitrogen excitation from the nearby carbon and oxygen excitations. With the WDS measurements, we used the same low voltage as for the EDS such that we could preferentially observe nitrogen, but there is no observable feature at the N-K$\alpha$ excitation energy, as shown in Fig. \ref{figEDSWDS}(b), which indicates that there is no nitrogen in the ambient pressure recovered sample.

\begin{figure}[h!]
\centering
\includegraphics[width=1\linewidth]{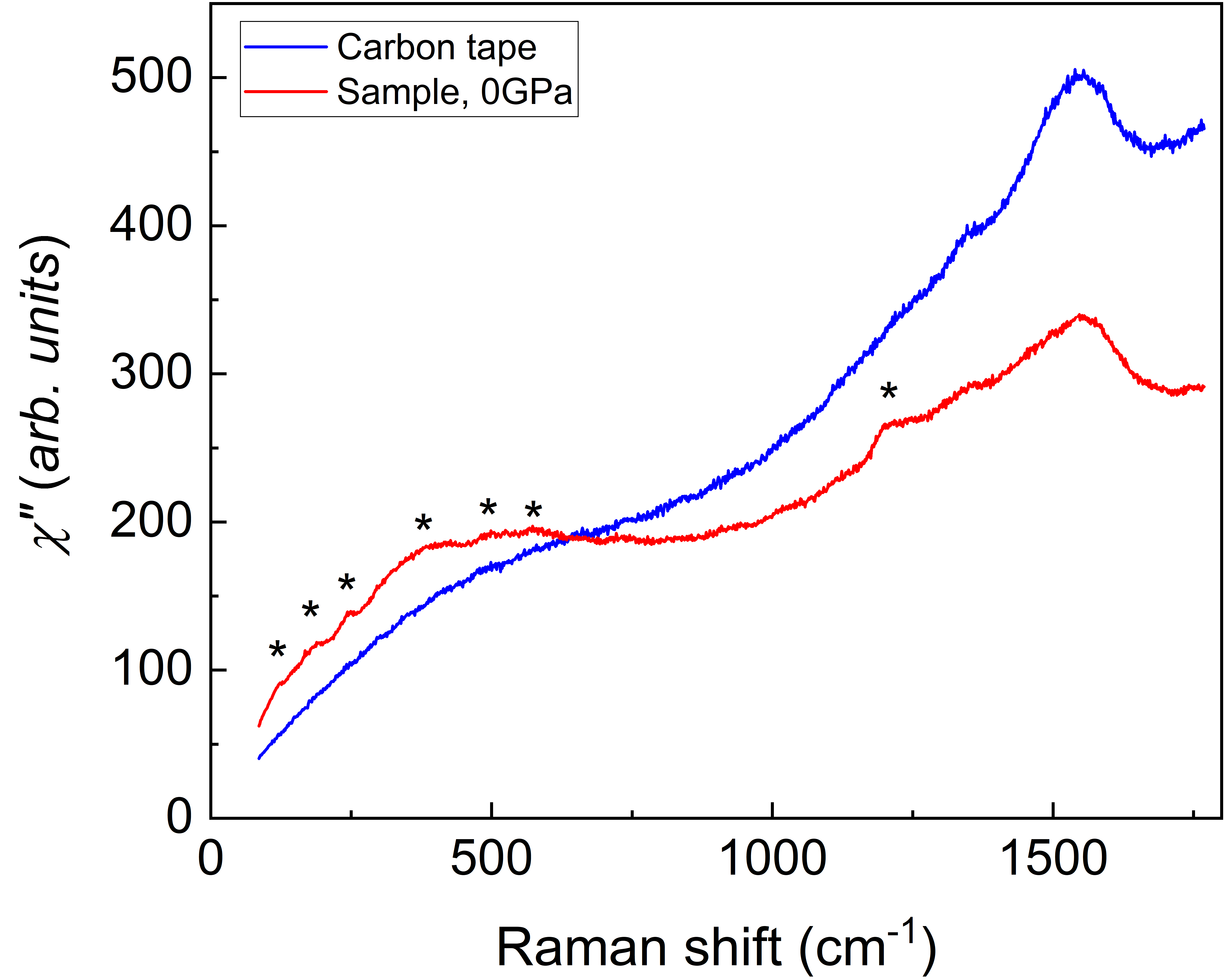}
\caption{Raman spectra of the recovered sample after measurements by EDS and WDS (12 days after opening). The sample is on carbon tape whose Raman spectrum is shown for comparison. The stars are the modes identified as coming from the sample. The subtracted spectrum is shown in Fig. \ref{RamanSI_FinalPhaseDecomp}(d).}
\label{RamanEDX}
\end{figure}

The EDS spectra were measured one day after opening the cell but before the Raman spectrum in Fig. \ref{figRaman2SI}(e), therefore it should still be somewhat representative of the cubic sample. The WDS was measured on day 2, but unfortunately, we do not know the state of the sample between 1\,day and 5\,days after opening. Fig. \ref{RamanEDX} shows the Raman spectrum of the recovered sample used for EDS measurements after 12\,days. We identify several peaks that do not originate from the carbon tape (the sample could not be removed from the tape due to its fragility). The background subtraction of this sample is shown in Fig. \ref{RamanSI_FinalPhaseDecomp}(d), where we clearly see the reappearance of the peak at 240\,\si{\centi \meter^{-1}} which was not present after 5\,days; however, this could be due to the large increase in background between day 1 and day 5 making the 240\,\si{\centi \meter^{-1}} peak immeasurable. By day 12, the background has decreased to a comparable value to day 1 and the 240\,\si{\centi \meter^{-1}} peak is observable again. The other peaks after 5\,days could be present albeit less intense and broader so overall they are less distinct.

\section{S5: Raman tables of structures}


Here we state the Raman and infrared excitations of various space groups for LuH$_x$ structures: cubic $Fm\overline{3}m$, trigonal $P\overline{3}c1$, hexagonal $P6_3cm$, $P6_3$, and $P6_3/mmc$, and finally cubic $Ia\overline{3}$ for Lu$_2$O$_3$ and $Ia\overline{3}$-type. The expected occupied Wyckoff positions for each space group are written in table \ref{tab3} which are then used to predict the number of excitations and their associated symmetries. The only clear expected selection rules concern the $A$ modes that are mainly Raman active in parallel polarisations, except in hexagonal symmetries where they are Raman-active in both. For the $Ia\overline{3}$-type phase that is proposed as the second phase at 1.9\,GPa, other Wyckoff positions should be occupied by hydrogen atoms. Unfortunately, due to the low scattering cross-section of hydrogen, we cannot determine which Wyckoff sites are occupied. Calculations would be helpful to determine the stoichiometry and the occupied sites.


\begin{table*}[]
\footnotesize
\begin{tabular}{c|c|c|c|c|c|c|c|c|c|c}
\hline \hline
       Space group & Point group & Lu$_{1}$ & Lu$_{2}$ & H$_{1}$ & H$_{2}$ & H$_{3}$& H$_{4}$& H$_{5}$ & Infrared-active & Raman-active    \\ \hline
$Fm\overline{3}m$ (LuH$_3$ \cite{Sun2023}) & $O_h$ & $4a$ & - & $8c$ & $4b$ & - & - & - & $2T_{1u}$ & $1T_{2g}$ \\ \hline
$Fm\overline{3}m$ (LuH$_2$ \cite{Sun2023}) & $O_h$ & $4a$ & - & $8c$ & - & - & - & - & $1T_{1u}$ & $1T_{2g}$ \\ \hline
$P\overline{3}c1$ (YH$_3$ \cite{Kierey2001}) & $D_{3d}$ & $6f$ & - & $2a$ & $4d$ & $12g$ & - & -  & $6A_{2u}\oplus11E_{u}$ & $5A_{1g}\oplus12E_{g}$  \\ \hline
$P6_3/mmc$ (YH$_3$ \cite{Kierey2001}) & $D_{6h}$ & $2c$ & - & $2a$ & $4f$ & - & - & - & $2A_{2u}\oplus2E_{1u}$ & $1A_{1g}\oplus1E_{1g}\oplus2E_{2g}$ \\ \hline
$P6_3cm$ (YH$_3$ \cite{Kierey2001}) & $C_{6v}$ & $6c$ & - & $6c$ & $6c$ & $4b$ & $2a$ & - & $7A_{1}\oplus11E_{1}$ & $7A_{1}\oplus11E_{1}\oplus12E_{2}$ \\ \hline
$P6_3$ (YH$_3$ \cite{Kierey2001}) & $C_{6}$ & $6c$ & - & $6c$ & $6c$ & $2b$ & $2b$ & $2a$ & $11A\oplus22E_1$ & $11A\oplus22E_1\oplus24E_2$   \\ \hline
$C2/m$ (YH$_3$ \cite{Yao2010}) & $C_{2h}$ & $2c$ & $4i$ & $8j$     & $4i$     & $2d$ & $4g$ & - & $7A_{u}\oplus11B_{u}$ & $8A_{g}\oplus7B_{g}$  \\ \hline \hline
$Ia\overline{3}$ ($Ia\overline{3}$-type) & $T_{h}$ & $8b$ & $24d$ & - & - & - & - & - & $7T_{u}$ & $1A_{g}\oplus2E_g\oplus5T_{g}$ \\ \hline
$Ia\overline{3}$ (Lu$_2$O$_3$ \cite{Jiang2010a}) & $T_{h}$ & $8b$ & $24d$ & $48e$ & - & - & - & - & $16T_{u}$ & $4A_{g}\oplus8E_g\oplus14T_{g}$ \\ \hline
$P6_3/mmc$ (pure Lu and Y \cite{Daou1974, Kierey2001}) & $D_{6h}$ & $2c$ & - & - & - & - & - & - & $1A_{2u}\oplus1E_{1u}$ & $1E_{2g}$ \\ \hline \hline
\end{tabular}
\caption{The total number of optical infrared and Raman-active modes for the given space groups with the occupied Wyckoff positions stated for various compounds.}
\label{tab3}
\end{table*}

\nocite{Pebler1962, FullProf, Prescher2015, Toby2013, Klotz2009, Bonnet1971, Eremets2011_1, Eremets2023_1, Sharma1980}

\end{document}